\documentclass[]{aa}  

\usepackage{graphicx}
\usepackage{txfonts}
\usepackage{color}
\usepackage[colorlinks=true, allcolors=blue]{hyperref}
\newcommand{\kms}{\ensuremath{\mathrm{km}\,\mathrm{s}^{-1}}}

\newcommand{\hi } {{\rm H}\,{\small\rm I}}
\newcommand{\ci } {[{\rm C}\,{\small\rm I}]}
\newcommand{\cii } {[{\rm C}\,{\small\rm II}]}
\newcommand{\nii } {[{\rm N}\,{\small\rm II}]}
\newcommand{\oii } {[{\rm O}\,{\small\rm II}]}
\newcommand{\oiii } {[{\rm O}\,{\small\rm III}]}
\newcommand{\bb } {\textsc{$^{\rm 3D}$Barolo}}

\begin{document}

   \title{Cold gas disks in main-sequence galaxies at cosmic noon: \\ Low turbulence, flat rotation curves, and disk-halo degeneracy}

   \titlerunning{Cold gas disks at cosmic noon}

   \author{Federico Lelli
          \inst{1}
          \and
          Zhi-Yu Zhang\inst{2,3}
          \and
          Thomas G. Bisbas\inst{4,5,6}
          \and
          Lingrui Lin\inst{2,3}
          \and
          Padelis Papadopoulos\inst{6,7}
          \and
          James M. Schombert\inst{8}
          \and
          Enrico Di Teodoro\inst{9}
          \and
          Antonino Marasco\inst{10}
          \and
          Stacy S. McGaugh\inst{11}
          }

   \institute{INAF, Arcetri Astrophysical Observatory, Largo E. Fermi 5, 50125, Florence, Italy; \email{federico.lelli@inaf.it}
    \and
    School of Astronomy and Space Science, Nanjing University, 163 Xianlin Avenue, Nanjing, Jiangsu, 210023, P.R. China
    \and
    Key Laboratory of Modern Astronomy and Astrophysics, Nanjing University, Ministry of Education, Nanjing 210023, P.R. China
    \and
    Research Center for Intelligent Computing Platforms, Zhejiang Laboratory, Hangzhou 311100, P.R. China
    \and
    Universit\"at zu K\"oln, I. Physikalisches Institut, Z\"ulpicher Str. 77, 50937 K\"oln, Germany
    \and
     Department of Physics, Section of Astrophysics, Astronomy and Mechanics, Aristotle University of Thessaloniki, GR-54124 Thessaloniki, Greece
    \and
    School of Physics and Astronomy, Cardiff University, Queens Buildings, The Parade, Cardiff CF24 3AA, UK
    \and
    Department of Physics, University of Oregon, 461 Willamette Hall, Eugene, OR 97403, USA
    \and
    Department of Physics \& Astronomy, University of Florence, via G. Sansone 1, 50019, Sesto Fiorentino, Italy
    \and
    INAF, Padova Astronomical Observatory, Vicolo Osservatorio 5, 35122, Padova, Italy
    \and
    Department of Astronomy, Case Western Reserve University, 10900 Euclid Avenue, Cleveland, OH 44106, USA
    } 
   \date{Received 30/09/2022; accepted 30/01/2023}

% \abstract{}{}{}{}{} 
% 5 {} token are mandatory
 
  \abstract
   {We study the dynamics of cold molecular gas in two main-sequence galaxies at cosmic noon (zC-488879 at $z\simeq1.47$ and zC-400569 at $z\simeq2.24$) using new high-resolution ALMA observations of multiple $^{12}$CO transitions. For zC-400569 we also reanalyze high-quality H$\alpha$ data from the SINS/zC-SINF survey. We find that (1)\,both galaxies have regularly rotating CO disks and their rotation curves are flat out to $\sim$8 kpc contrary to previous results pointing to outer declines in the rotation speed $V_{\rm rot}$; (2)\,the intrinsic velocity dispersions are low ($\sigma_{\rm CO}\lesssim15$ \kms\, for CO and $\sigma_{\rm H\alpha}\lesssim37$ \kms\, for H$\alpha$) and imply $V_{\rm rot}/\sigma_{\rm CO}\gtrsim17-22$ yielding no significant pressure support; (3)\,mass models using HST images display a severe disk-halo degeneracy, that is models with inner baryon dominance and models with ``cuspy'' dark matter halos can fit the rotation curves equally well due to the uncertainties on stellar and gas masses; and (4)\,Milgromian dynamics (MOND) can successfully fit the rotation curves with the same acceleration scale $a_0$ measured at $z\simeq0$. The question of the amount and distribution of dark matter in high-$z$ galaxies remains unsettled due to the limited spatial extent of the available kinematic data; we discuss the suitability of various emission lines to trace extended rotation curves at high $z$. Nevertheless, the properties of these two high-$z$ galaxies (high $V_{\rm rot}/\sigma_{\rm V}$ ratios, inner rotation curve shapes, bulge-to-total mass ratios) are remarkably similar to those of massive spirals at $z\simeq0$, suggesting weak dynamical evolution over more than 10 Gyr of the Universe's lifetime.}

   \keywords{dark matter -- galaxies: evolution -- galaxies: formation -- galaxies: high-redshift -- galaxies: kinematics and dynamics}

   \maketitle
%
%-------------------------------------------------------------------

\section{Introduction}\label{sec:intro}

During the past decades, there has been outstanding progress in studying the internal dynamics of high-$z$ galaxies. Near-infrared (NIR) spectroscopy with integral field units (IFUs) allowed for the kinematics of warm ($T\simeq10^4$ K) ionized gas to be traced using the H$\alpha$ emission line at $z\simeq1.0-2.5$ \citep[e.g.,][]{Forster2009, Wisnioski2015, Stott2016} and the \oiii$\lambda$5007 \AA\ line up to $z\simeq3.5$ \citep[e.g.,][]{Gnerucci2011, Turner2017a}. Radio and submillimeter observations with the Jansky Very Large Array (JVLA) and the NOrthern Extended Millimeter Array (NOEMA) allowed for the kinematics of cold neutral gas ($T\simeq10-100$ K) to be traced using CO transitions at $z\simeq1-4$ \citep[e.g.,][]{Hodge2012, Ubler2018}. Moreover, the Atacama Large Millimeter Array (ALMA) made it possible to study gas dynamics using \ci\ lines at $z\simeq1-3$ \citep{Lelli2018,  Dye2022, Gururajan2022}, the \cii\ line at $z\simeq4-7$ \citep{DeBreuck2014, Jones2017, Smit2018}, and high-$J$ CO lines \citep{Tadaki2017, Talia2018}.

The first IFU surveys of massive galaxies ($M_{\star} \gtrsim 10^{10} M_{\odot}$) at $z\simeq1-3$ suggested that about one-third of star-forming galaxies were rotation-dominated disks, one-third were dispersion-dominated objects, and another third were merging systems \citep{Forster2009, Gnerucci2011}. Subsequent observations, however, led to a drastically different picture: it is now clear that at least 80$\%$ of star-forming galaxies at $z\simeq1-3$ have rotating gas disks, whereas dispersion-dominated and merging systems constitute a minority of the star-forming population \citep{Wisnioski2015, Wisnioski2019, Stott2016}. The reasons for such a changing view are complex, but an important role has been played by the limited spatial resolution and the resulting beam-smearing effects \citep{Warner1973, Bosma1978, Begeman1989}. When a rotating disk is spatially resolved with only a few resolution elements, different line-of-sight velocity projections are flux-averaged within the resolution element creating two main observational effects: (1) the line profiles are artificially broadened, so a rotation-supported disk may appear as a dispersion-dominated object when observed at low spatial resolution, and (2) the line profiles become asymmetric, typically with long tails of emission toward the systemic velocity, so the gas kinematics may appear more complex than they really are. These two observational effects can then lead to an artificially high fraction of dispersion-dominated and merging systems with respect to rotation-dominated ones. The same issues may occur in low-resolution \cii\ surveys of galaxies at $z\simeq4-7$ \citep{LeFevre2020, Jones2021, Neeleman2021}.

While there is now overall agreement about the existence of rotating disks at high $z$, their kinematic properties remain debated. Kinematic studies of warm ionized gas led to the common view that high-$z$ disks are more turbulent than their local analogs (\citealt{Forster2009, Lehnert2009, Gnerucci2011}). The gas velocity dispersion ($\sigma_{\rm V}$) is thought to increase systematically with $z$, while the degree of rotation support ($V_{\rm rot}/\sigma_{\rm V}$) decreases \citep{Wisnioski2015, Stott2016, Ubler2019}. Low values of $V_{\rm rot}/\sigma_{\rm V}$ would imply that the gas disk is not fully supported by rotation: if one aims to trace the circular velocity of a test particle in the equilibrium gravitational potential, the pressure support would need to be taken into account (the so-called asymmetric-drift correction). For example, \citet{Genzel2017} and \citet{Lang2017} found that H$\alpha$ rotation curves decline in the outer region steeper than the Newtonian expectation for a thin disk, and argued that such a super-Keplerian decline is due to pressure gradients from strong, turbulent gas motions.

Kinematic studies of cold neutral gas from ALMA are painting a different picture. High-resolution observations of CO, \ci\ and \cii\ lines revealed the existence of gas disks with low turbulence and high $V_{\rm rot}/\sigma$ values at $z\simeq2-7$ \citep{Lelli2018, Lelli2021, Kaasinen2020, Fraternali2021, Xiao2022, Posses2023}, so the observed rotation curves do not require corrections for pressure support. Dynamically cold gas disks have also been found in gravitationally lensed galaxies, in which the effective spatial resolution is exceptionally high ($\sim$100$-$500 pc) due to the magnification effect \citep{DiTeodoro2018, Rizzo2020, Rizzo2021, Dye2022}. 

The cause of the disagreement between optical and submillimeter studies may be twofold. On the one hand, beam-smearing effects and their modeling may again play a role: even when a rotating disk is clearly recognized, the finite spatial resolution has nontrivial effects in assessing its kinematic properties. The artificial broadening of the emission line can lead to severe overestimates of the gas velocity dispersion, while the artificial skewness can lead to severe underestimates of the rotation velocity \citep[e.g.,][]{Swaters2009, DiTeodoro2015, DiTeodoro2016}. For example, one may derive a slowly rising rotation curve even though the intrinsic rotation curve rises steeply \citep[e.g.,][]{Lelli2010}. In addition to getting observations with the highest possible spatial resolution, the best approach to account for beam-smearing effects is modeling the three-dimensional (3D) emission-line datacube in a nonparametric fashion \citep[e.g.,][]{Corbelli1997, Sicking1997, Swaters1999, Fraternali2001, Fraternali2002, Gentile2003, Gentile2004, Gentile2007, Jozsa2007, Swaters2009, Lelli2010, Lelli2012a, Lelli2012b, Lelli2014, DiTeodoro2015}.

On the other hand, the different gas phases probed by different emission lines may also play a role. At $z\simeq0$, ionized gas disks tend to display higher velocity dispersions and more complex noncircular motions than atomic and molecular gas disks \citep[e.g.,][]{Lelli2022}. A similar behavior may occur at high $z$. The velocity dispersions from CO lines, indeed, appear systematically smaller than those from the H$\alpha$ line \citep{Ubler2019}. In turn, H$\alpha$ velocity dispersions seem to be smaller than those from the \oii\ line and stellar absorption lines \citep{Ubler2022}.

%\subsection{This work: Two birds with one stone}
\begin{figure*}
\centering
\includegraphics[width=0.9\textwidth]{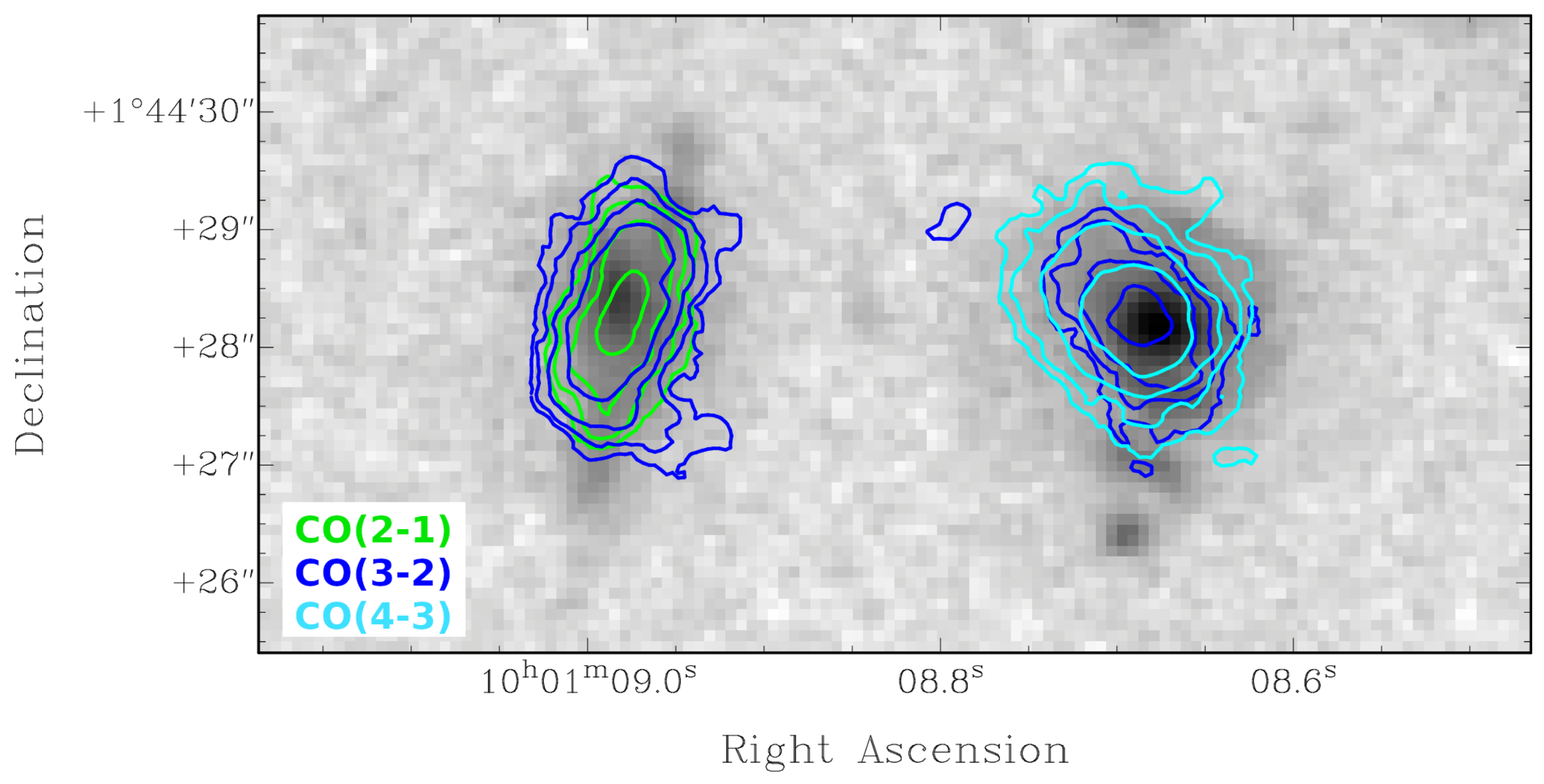}
\caption{The galaxies zC-488879 at $z\simeq1.47$ (left) and zC-400569 at $z\simeq2.24$ (right). The HST/WFC3 image in the NIR F160W filter (gray scale) is overlaid with CO($2-1$) emission (green contours), CO($3-2$) emission (blue contours), and CO($4-3$) emission (cyan contours). Contours are at (3, 6, 12, 24) $\sigma_{\rm map}$ with the values of 3$\sigma_{\rm map}$ given in Table\,\ref{tab:cubes}. The band 4 observations have lower quality than the band 3 ones, so the different spatial extent of CO($3-2$) and CO($4-3$) emission in zC-400569 may be due to sensitivity rather than being physical.}
\label{fig:TwoBirds}
\end{figure*}
To shed new light on galaxy dynamics at cosmic noon, we present an in-depth study of two galaxies with kinematic and photometric data of the highest quality. We used the Atacama Large Millimeter/submillimeter Array (ALMA) to obtain high-resolution data of the CO($3-2$) and CO($4-3$) lines for a main-sequence galaxy at $z\simeq2.24$: zC-400569 \citep{Mancini2011} also known as COSMOS\,0488950 \citep{Capak2007, Mobasher2007} or COSMOS2015\,0300906 \citep{Laigle2016}. This galaxy is one of the best observed object from the SINS IFU survey \citep{Forster2009, Forster2018, Genzel2017}. Luckily, next to zC-400569, there is another main-sequence galaxy at $z=1.47$ (COSMOS\,0488879 or COSMOS2015\,0301356) so we set up our ALMA observations to simultaneously target the CO($2-1$) and CO($3-2$) lines of this second object (see Fig.\,\ref{fig:TwoBirds}). For brevity, we refer to this second galaxy as zC-488879. Essentially, we got two birds with one stone.

\citet{Liu2019} provide the stellar mass ($M_\star$) and star-formation rate (SFRs) of both galaxies by fitting their spectral energy distributions (SEDs), combining optical and NIR photometry from COSMOS with the ALMA submillimeter continuum. According to their work, zC-400569 has $M_\star=2 \times 10^{11}$ M$_\odot$ and $\rm{SFR = 81}$ M$_\odot$ yr$^{-1}$ while zC-488879 has $M_\star=5 \times 10^{10}$ M$_\odot$ and $\rm{SFR = 115}$ M$_\odot$ yr$^{-1}$. Thus, both galaxies lie on the star-formation main sequence within the observed scatter \citep[cf. with][]{Schreiber2015}.

\begin{table*}
\caption[]{Properties of ALMA data. All cubes have channel width of 15 \kms and velocity resolution of 30 \kms after Hanning smoothing.}
\centering
\footnotesize
\label{tab:cubes}
\renewcommand{\arraystretch}{1.3}
\begin{tabular}{ccccccccc}
\hline
Galaxy & Redshift & Line &  Beam & Beam & Beam PA & $\sigma_{\rm cube}$ & 3$\sigma_{
\rm map}$ & Flux\\
       &      & & arcsec$\times$arcsec & kpc$\times$kpc & degree & mJy/beam & mJy/beam \kms & Jy \kms\\
\hline
zC-400569 & 2.23999 & CO(3-2) & $0.41\times0.34$ & $3.5\times2.9$ & $-83.9$ & 0.09 & 11.3 & 0.50$^{+0.09}_{-0.08}$\\
& & CO(4-3) & $0.54\times 0.46$ & $4.6\times3.9$ & $-84.2$& 0.07 & 13.0 & 0.76$^{+0.08}_{-0.08}$\\
& & \ci(1-0) & $0.58\times0.49$ & $ 4.9 \times 4.1$ & $-82.1$ & 0.07 & 10.0 & 0.51$^{+0.04}_{-0.06}$\\
zC-488879 & 1.46997 & CO(2-1) & $0.48\times0.40$ &  $4.2 \times 3.5$ & $-84.8$ & 0.08 & 15.1 & 0.56$^{+0.04}_{-0.05}$\\ 
& & CO(3-2) & $0.46\times0.39$ & $4.0 \times 3.4$ & $-82.8$ & 0.09 & 10.3 & 1.0$^{+0.07}_{-0.09}$\\
\hline
\end{tabular}
\end{table*}

Throughout this paper, we assume a flat $\Lambda$ cold dark matter ($\Lambda$CDM) cosmology with $H_{0}= 67.4$ km s$^{-1}$ Mpc$^{-1}$, $\Omega_{\rm m} = 0.315$, and $\Omega_{\Lambda} = 0.685$ \citep{Planck2018}. In this cosmology, 1 arcsec corresponds to 8.45 kpc at $z= 2.24$ and 8.68 kpc at $z=1.47$. The age of the Universe and the light travel time are, respectively, 2.9 Gyr and 10.9 Gyr at $z= 2.24$, and 4.3 Gyr and 9.5 Gyr at $z=1.47$.
%--------------------------------------------------------------------

\section{Data analysis}

\subsection{ALMA observations}\label{sec:ALMA}

ALMA observations were obtained through two different projects (2017.1.01020.S and 2019.1.00862.S; PI: T. Bisbas): the first one provides low-resolution observations in band 4, while the second one is a high-resolution follow-up in both band 4 and band 3. The two galaxies (zC-400569 and zC-488879) were simultaneously observed in the same field of view, pointing toward the center of zC-400569. In both bands, we used a mixed spectral setup with four spectral windows (SPWs) with a bandwidth of 1.875 GHz each. For band-3 observations, one SPW was used to cover the CO($2-1$) line of zC-488879, another one for the CO($3-2$) line of zC-400569, and the remaining two SPWs for the 3-mm continuum. For band-4 observations, the four SPWs cover the CO($3-2$) line of zC-488879, the CO($4-3$) line of zC-400569, the \ci($1-0$) line of zC-400569, and the 2-mm continuum. All line SPWs were covered with 1920 channels giving a channel width of 4$-$5 \kms, while the continuum SPWs were covered with 128 channels giving a channel width of 60$-$100 \kms.

In project 2017.1.01020.S, band-4 observations used the nominal C43-3 array configuration with 41 to 45 antennas, providing minimum and maximum baselines of 14 and 740 m, respectively. Three execution blocks were obtained in May 2018 giving a total on-source integration time of about 2.5 hours. In project 2019.1.00862.S, band-3 and band-4 observations used the nominal C43-6 configuration with 40 to 45 antennas, providing minimum and maximum baselines of 14 and 3638 m, respectively. Five executions blocks were obtained in band 4 in May 2021, while six execution blocks were obtained in band 3 between May and July 2021. Unfortunately, the execution blocks taken on 11 May 2021 (band 4) and 26 May 2021 (band 3) did not reach the expected quality and are therefore not used in this study. The final on-source integration time is $\sim$4 hours in band 3 and $\sim$3 hours in band 4.

The data reduction was performed using the Common Astronomy Software Applications (\textsc{Casa}) package \citep{McMullin2007, CASA2022}. The Fourier-plane data were flagged and calibrated using the appropriate pipeline version provided by the ALMA team, which varies for different observing runs. The imaging was performed with the \texttt{tclean} task in \textsc{Casa} (version 6.1.2.7) using a Briggs' robust parameter of 1.5 and interactive cleaning. Emission-line cubes were derived using a channel width of $\sim$15 \kms\, to have adequate signal-to-noise ratio (S/N). In the individual SPWs, the continuum emission is undetected, so no continuum subtraction was performed. For our kinematic analysis (Sect.\,\ref{sec:gaskin}), we use cubes without primary-beam correction, so the noise structure is uniform and well behaved. A primary-beam map is used to correct moment-one maps and calculate total fluxes (as we describe below) but the flux correction is entirely negligible for zC-400569 and only 1-2$\%$ for zC-488879 because both galaxies are well within the half-power beam-width (HPBW) of $41''$ in band 3 and $62''$ in band 4.

After imaging, the emission-line cubes were analyzed using the \bb\ software \citep{DiTeodoro2015} (version 1.6). To further boost the S/N, the cubes were Hanning smoothed over three channels using the \texttt{Smoothspec} task, giving a velocity resolution of 30 km s$^{-1}$ and the rms noise ($\sigma_{\rm cube}$) listed in Table\,\ref{tab:cubes}. Subsequently, moment maps were obtained with the \texttt{Makemask} task, considering the signal inside a Boolean mask that was created with the \texttt{Smooth \& Search} task. When using a mask, the noise in the moment-zero map ($\sigma_{\rm map}$) varies from pixel to pixel. Following \citet{Verheijen2001} and \citet{Lelli2014c}, we build a S/N map considering channel dependencies, then we computed a pseudo 3$\sigma_{\rm map}$ value taking the median intensity of pixels with S/N between 2.9 and 3.1. Total line fluxes are measured summing pixels with $\rm{S/N>3}$ in the moment-zero maps; uncertainties are estimated repeating the sum for pixels with $\rm{S/N>2}$ and $\rm{S/N>4}$. Emission-line maps are discussed in Sect.\,\ref{sec:gasdist}, but we anticipate that our kinematic analysis in Sect.\,\ref{sec:gaskin} fits directly the 3D cubes, modeling the effects of beam smearing on the emission-line profiles at each spatial location. Table\,\ref{tab:cubes} summarizes the properties of the ALMA data.

\begin{figure*}
\centering
\includegraphics[width=0.495\textwidth]{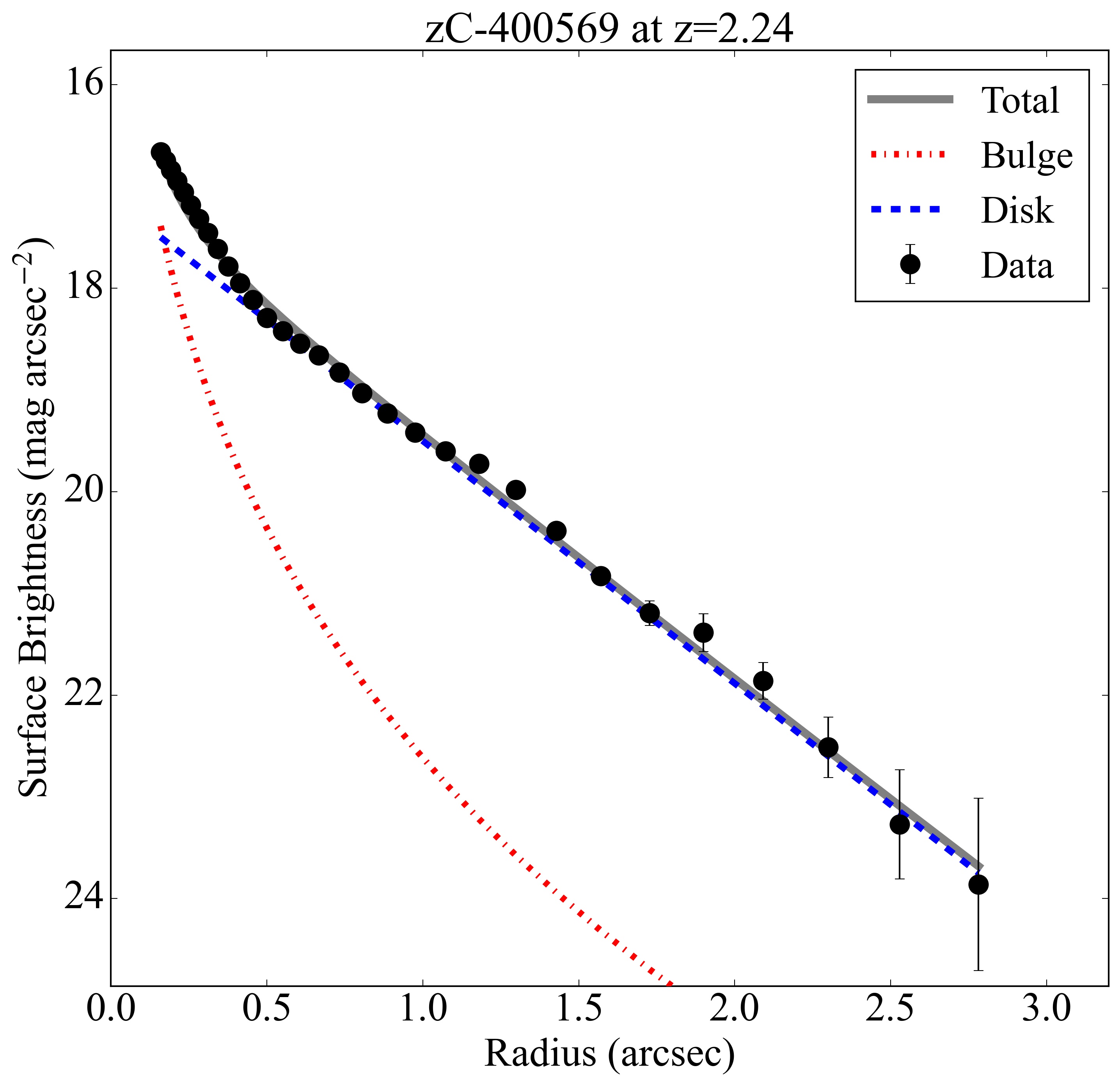}
\includegraphics[width=0.495\textwidth]{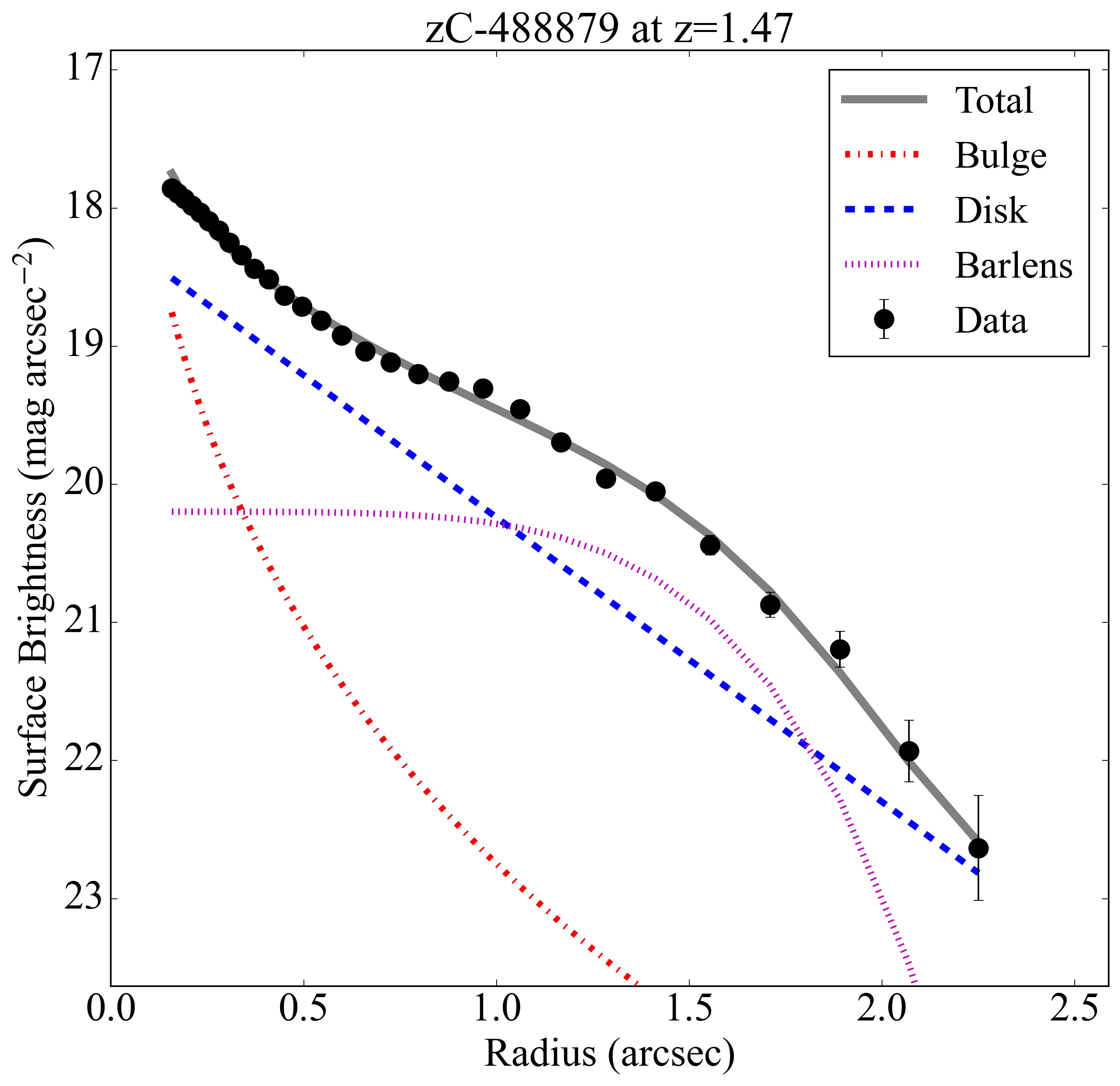}
\caption{Surface brightness profiles of zC-400569 (left panel) and zC-488879 (right panel) from the NIR F160W filter of HST/WPC3. The observed profile (black dots) is fitted with a multicomponent model (gray solid line) considering a De Vaucouleurs' bulge (red dash-dotted line), an exponential disk (blue dashed line), and a barlens component (magenta dotted line, if present).}
\label{fig:SBprof}
\end{figure*}
\subsection{SINFONI data}\label{sec:SINF}

The SINFONI data of zC-400569 were obtained from the archive of the SINS/zC-SINF Adapative Optics (AO) survey\footnote{\url{https://www.mpe.mpg.de/ir/SINS/SINS-zcSINF-data}} \citep{Forster2018}. We first add the WCS coordinate system to the cube and convert the third axis unit from $\mu$m to velocity assuming the H$\alpha$ redshift of 2.2405 \citep{Forster2018}. To align the cube with the north direction, we perform a rotation of 30 degrees counterclockwise at the kinematic center of the galaxy and resample the data with a pixel size of 0.05$''$, the same as its original value. Finally, we checked the astrometry by overlaying the moment-zero map from the SINFONI cube with images from the Hubble Space Telescope (HST), which are discussed in Sect.\,\ref{sec:HST}. The full-width half-maximum (FWHM) of the point spread function (PSF) of the SINFONI cube is $\sim$0.32$''$, while the FWHM of the line spread function is $\Delta_{\rm V}\simeq87$ \kms\ ($\sigma_{\rm inst}
\simeq \Delta_{\rm V}/2.35 \simeq 37$ \kms) around the H$\alpha$ line \citep{Forster2018}.

At the spatial location of the central galaxy, continuum emission is detected. The continuum emission is noisy and no stellar absorption line is detected. To estimate the continuum flux and subtract it from the SINFONI cube, we fit a first order polynomial to the line-free channels, excluding spectral regions with H$\alpha$ and \nii\ emission. We explored the use of higher-order polynomials but they did not significantly improved the continuum subtraction. After continuum subtraction, we checked that the flux ratios of the two \nii\ lines are consistent with the theoretical value within the errors. Finally, we produced a sub-cube that covers only the H$\alpha$ emission region. This continuum-subtracted H$\alpha$-only cube will be used in the rest of our analysis.

\subsection{HST data}\label{sec:HST}

Both zC-400569 and zC-488879 have been detected by HST in various filters. Since we aim to model the stellar mass distribution of these galaxies, we analyzed the reddest HST images available in the Hubble Legacy Archive\footnote{\url{https://hla.stsci.edu/}}, which were taken with the NIR F160W and F110W filters of the Wide Field Camera 3 (WFC3; Project ID 12578). The FWHM of the PSF in the F110W and F160 images are 0.13" and 0.15", respectively. At the redshift of zC-400569 ($z\simeq2.24$), the F110W filter is comparable to rest-frame Cousins $U$ band, while the F160W filer is in-between rest-frame $V$ and $B$ bands. At the redshift of zC-488879 ($z\simeq1.47$), the F110W and F160W filters are comparable to rest-frame $B$ and $R$ bands, respectively.

The F160W image is shown in Figure\,\ref{fig:TwoBirds} overlaid with the CO moment-zero maps derived in Sect.\,\ref{sec:ALMA}. Overall, the HST astrometry is in good agreement with the ALMA one. For zC-400569 the peak emission in the HST image is slightly to the southwest of the centroid of the innermost CO contour, so one may have the visual impression that the HST astrometry needs to be rectified by 2-3 pixels to the top-left direction. On the other hand, for zC-488879 the peak emission in the HST image is slightly to the northeast of the centroid of the innermost CO contour, so one may have the visual impression that the HST astrometry needs to be rectified by 2-3 pixels in the opposite direction. Distortions in the HST camera cannot be so large, so the minor off-sets between stellar and CO distributions are not driven by technical issues and we apply no corrections to the HST astrometry.

We performed standard surface photometry using the \textsc{Archangel} software\footnote{\url{http://abyss.uoregon.edu/~js/archangel/}} \citep{Schombert2011}. \textsc{Archangel} has been widely used on local galaxies of all morphological types, from giant ellipticals to low-surface brightness dwarf galaxies, using images from various telescopes operating from NIR to ultra-violet bands \citep{Tully2009, Tully2013, Schombert2011, Schombert2016, Schombert2018, Schombert2012, Beygu2017, Greene2019}. In short, after frame cleaning and sky determination, \textsc{Archangel} fits elliptical isophotes to the images and derives azimuthally averaged surface brightness profiles (see previous references for details). The resulting profiles are shown in Figure\,\ref{fig:SBprof} for the F160W filter. The profiles from the F110W filter show a similar shape but they have larger errors, so they will not be used in the rest of this paper.

\begin{table}
\caption[]{Best-fit results from fitting the surface brightness profiles with a multicomponent galaxy model (see Fig.\,\ref{fig:SBprof}). Surface brightnesses are measured quantities with no cosmological corrections.}
\centering
\footnotesize
\label{tab:photo}
\renewcommand{\arraystretch}{1.3}
\begin{tabular}{lcc}
\hline
Parameter & zC-400569 & zC-488879 \\
\hline
$\mu_{\rm e,\,bul}$ (mag arcsec$^{-2}$) & 8.4$\pm$2.7 & 11.9$\pm$4.9\\ 
$R_{\rm e,\,bul}$ (arcsec)  & 0.12$\pm$0.13& 0.35$\pm$1.1\\
$\mu_{\rm e,\,disk}$ (mag arcsec$^{-2}$) &      17.1$\pm$0.1 & 18.2$\pm$0.4\\ 
$R_{\rm e,\,disk}$ (arcsec)  & 0.76$\pm$0.01 & 0.88$\pm$0.05 \\
$\mu_{\rm e,\,lens}$ (mag arcsec$^{-2}$) & ...     & 20.2$\pm$0.2\\ 
$R_{\rm e,\,lens}$ (arcsec)  & ... & 1.10$\pm$0.03\\
\hline
\end{tabular}
\end{table}
For zC-400569, the surface brightness profile is well fitted by a classic bulge plus disk model, in which the bulge is described by a \citet{DeVauc1948} profile and the disk by an exponential profile \citep{Freeman1970}. The model has four free parameters: the effective radius and effective surface brightness of the bulge ($R_{\rm e,\,bul}$ and $\mu_{\rm e, \,bul}$) and those of the disk ($R_{\rm e,\,disk}$ and $\mu_{\rm e,\,disk}$). For zC-400569, the same model does not provide satisfactory results because the surface brightness profile displays a curvature at $1''<R<2''$ that is reminiscent of ``lenses'' and ``barlenses'' in disk galaxies at $z\simeq0$ \citep{Laurikainen2011, Athanassoula2015}. Thus, we add a lens-like component using a \citet{Sersic1963} profile with $n=0.2$ and two further free parameters: $R_{\rm e,\,lens}$ and $\mu_{\rm e,\,lens}$. Table\,\ref{tab:photo} provides the results of our fits, which were performed using the orthogonal-distance-regression algorithm in the \texttt{SciPy} package of \textsc{Python} \citep{2020SciPy}. For both galaxies the bulge parameters are very uncertain (especially $R_{\rm e,\,bul}$) because most of the bulge light is contained within the PSF ($\sim$0.15$''$ or $\sim$1.3 kpc for the adopted cosmology). To properly calculate the stellar gravitational potential, therefore, the bulge profile needs to be extrapolated below the HST spatial resolution (see Sect.\,\ref{sec:massmodels}).

\section{Gas distribution \& kinematics}\label{sec:gasdist}

\subsection{The galaxy zC-400569 at $z\simeq2.4$}

\begin{figure}
\centering
\includegraphics[width=0.48\textwidth]{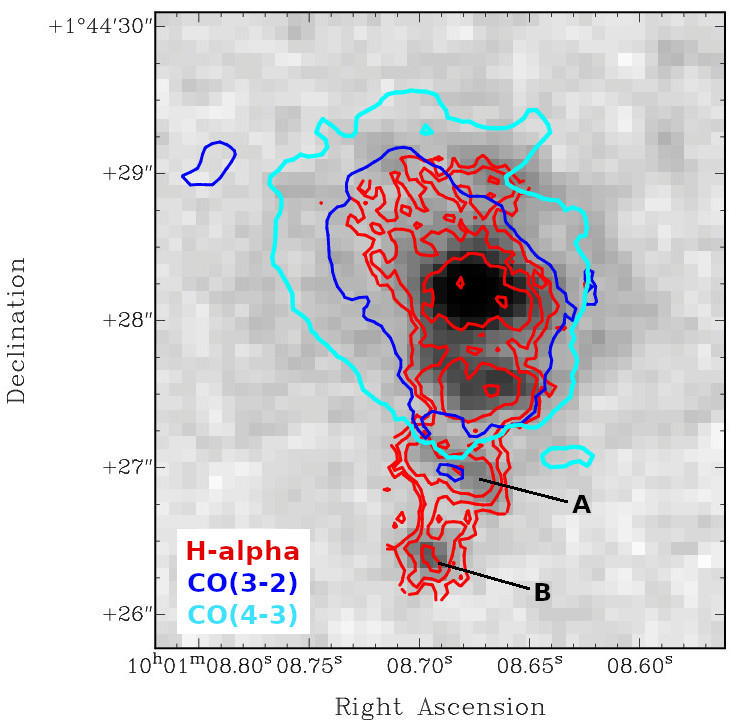}
\caption{Ionized and molecular gas in zC-400569 at $z\simeq2.24$. The HST/WFC3 image in the NIR F160W filter (gray colorscale) is overlaid with H$\alpha$ emission (red contours), CO($3-2$) emission (blue contours), and CO($4-3$) emission (cyan contours). Arrows indicate two possible dwarf companions (``A'' and ``B''), which are bright in H$\alpha$ but undetected in CO and \nii\ lines, suggesting low metallicities.}
\label{fig:SINFONI}
\end{figure}

Figure\,\ref{fig:SINFONI} compares the distributions of molecular and ionized gas in zC-400569 at $z\simeq2.24$. The H$\alpha$ emission display a ``clumpy'' morphology with a northsouth orientation. The two brightest H$\alpha$ components to the north are encompassed by the CO emission, whereas the two weakest H$\alpha$ components to the south are undetected in any CO line. In these two components, which we name ``A'' and ``B'', the \nii\ lines are undetected too \citep[in agreement with][]{Genzel2017}. These two ``clumps'' are also visible in the HST images. Most likely, they are low-mass, low-metallicity galaxies that are interacting with the main one, as previously suggested by \citet{Genzel2017}.

\begin{figure*}
\centering
\includegraphics[width=0.95\textwidth]{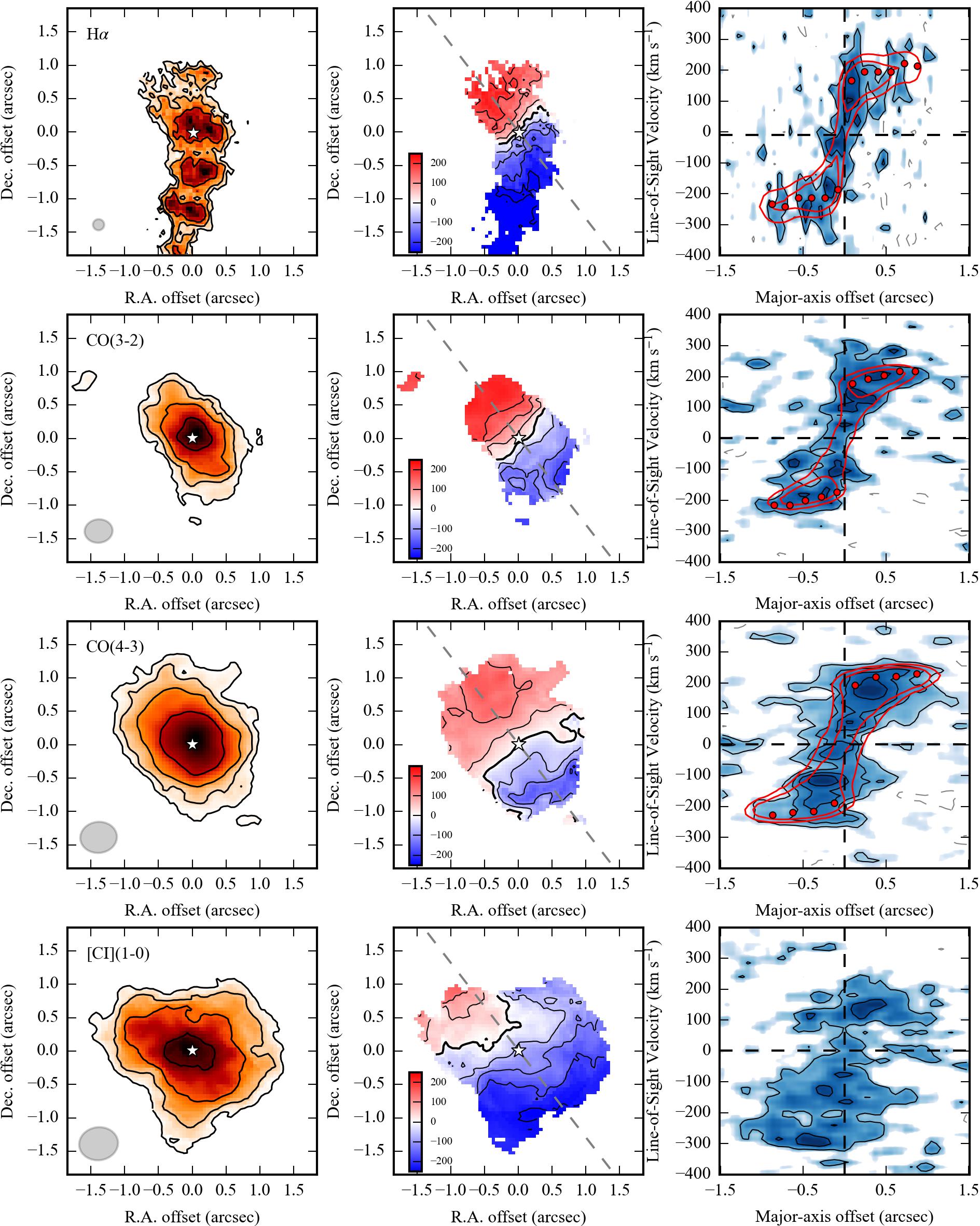}
\caption{Overview of zC-400569 at $z\simeq2.24$. \textit{Left panels:}
Total intensity maps for different emission lines, indicated in the top-left corner. Contours are at (3, 6, 12, 24) $\sigma_{\rm map}$ with the 3$\sigma_{\rm map}$ values given in Table\,\ref{tab:cubes}. The white star shows the kinematic center. The gray ellipse to the bottom-left corner represents the PSF. \textit{Middle panels:} Velocity fields. The thick contour corresponds to the systematic velocity (set to zero) and the thin contours are in steps of $\pm$60 \kms. The white star shows the kinematic center and the dashed line shows the disk major axis. \textit{Right panels:} Position-velocity diagrams along the disk major axis. The blue colorscale shows the observed gas emission. Black contours range from 2$\sigma_{\rm cube}$ to 8$\sigma_{\rm cube}$ in steps of 2$\sigma_{\rm cube}$. Red contour (at 2$\sigma_{\rm cube}$ and 4$\sigma_{\rm cube}$) shows the best-fit 3D kinematic model from \bb. The red dots show the best-fit sky-projected rotation curve. The vertical and horizontal dashed lines correspond to the galaxy center and systemic velocity, respectively.}
\label{fig:Mosaic1}
\end{figure*}

\begin{figure*}
\centering
\includegraphics[width=0.95\textwidth]{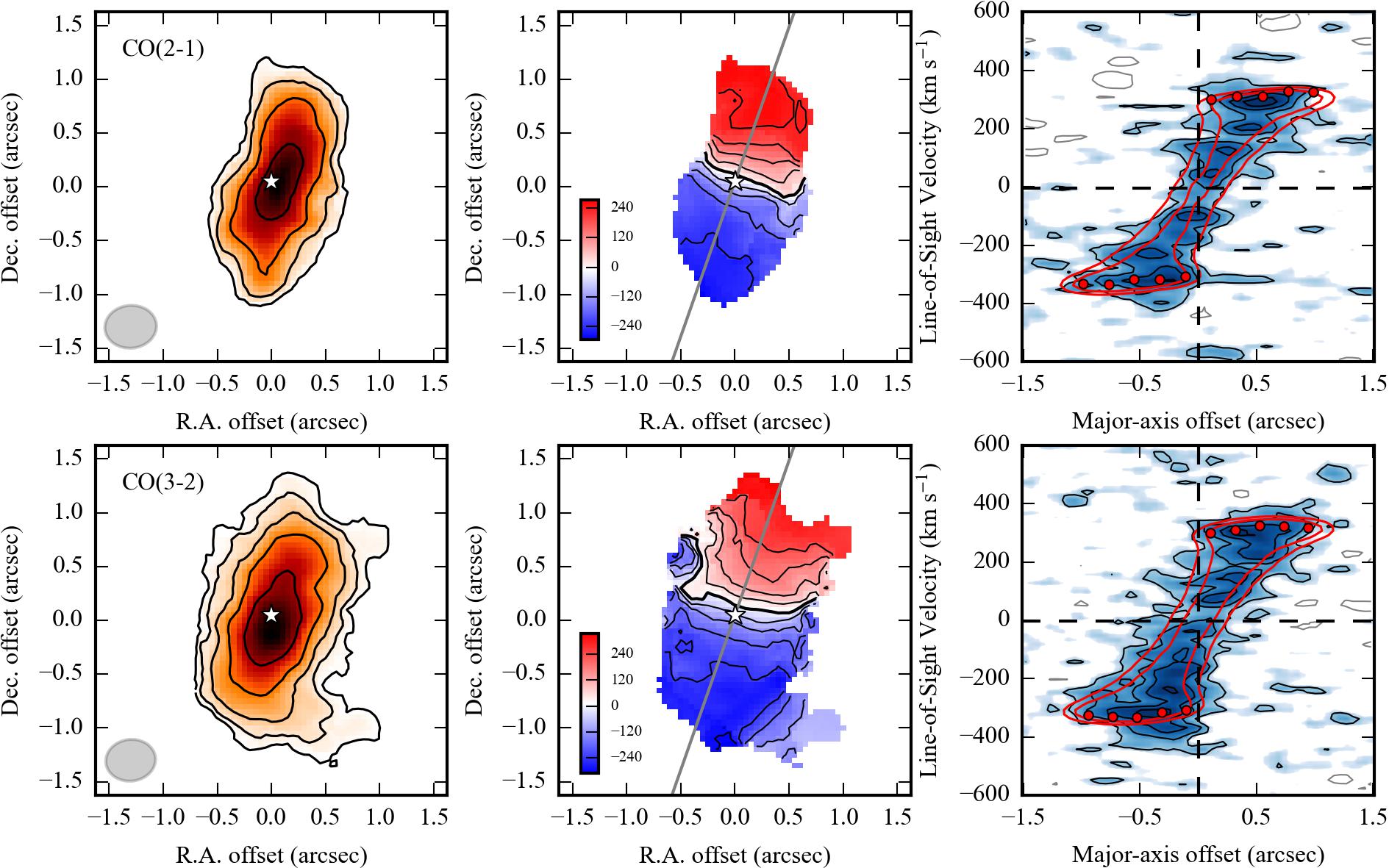}
\caption{Overview of zC-488879 at $z\simeq1.47$. For the description of the panels, see the caption of Fig.\,\ref{fig:Mosaic1}.}
\label{fig:Mosaic2}
\end{figure*}

Figure\,\ref{fig:Mosaic1} provides an overview of all emission line data, showing total intensity maps, line-of-sight velocity maps, and position-velocity (PV) diagrams along the galaxy major axis (see Sect.\,\ref{sec:gaskin} for details). The CO distribution appears smoother than the H$\alpha$ distribution, but this may possibly be an effect of the lower spatial resolution. Any giant clump in molecular gas must be significantly smaller than the ALMA beam ($3-4$ kpc). The spatial distributions of CO($4-3$) and CO($3-2$) emission are in good agreement.

In terms of kinematics, CO and H$\alpha$ emissions form regularly rotating disks, as shown by velocity fields and PV diagrams. This implies that stellar feedback does not destroy the overall kinematic regularity of the gas disk, similarly to the situation in local spirals where stellar feedback mostly drives a kpc-scale gas circulation via galactic fountains \citep[e.g.,][]{Marasco2019, LMF2021}. The possible interaction with components A and B appears to have a negligible effect on the overall regularity of the main disk, albeit some minor kinematic disturbances are visible toward the southeast of the velocity fields. We stress that details in the velocity fields should be interpreted with caution because of beam smearing effects; the gas kinematics is studied in Sect.\,\ref{sec:gaskin} using 3D models.

The \ci($1-0$) emission seems to trace the same rotating disk of the other lines but the major axis PV diagram is asymmetric and noisy. Moreover, the \ci\ systematic velocity appears offset with respect to that from the H$\alpha$ and CO lines. We were not able to perform a detailed kinematic analysis of the \ci\ data, so we use them solely as a total H$_2$ mass tracer in Sect.\,\ref{sec:onlybaryons}.

\subsection{The galaxy zC-488879 at $z\simeq1.47$}

Figure\,\ref{fig:Mosaic2} provides an overview of the emission-line data of zC-488879 at $z\simeq1.47$. Similarly to zC-400569, the CO distribution is smooth and does not show the large clumps that are often seen in rest-frame UV and H$\alpha$ images of galaxies at cosmic noon \citep[e.g.,][]{Zanella2019}. The spatial distributions of CO($3-2$) and CO($2-1$) emission are in overall agreement, but the CO($2-1$) emission peaks near the galaxy center while the CO($3-2$) emission peaks toward the south. This may be possibly related to varying CO line excitation conditions in the galaxy. 

Both CO($2-1$) and CO($3-2$) lines indicate regularly rotating disks, as shown by velocity maps and PV diagrams. The CO($3-2$) emission has an extension to the southwest, which is kinematically connected with the main disk and display a small velocity gradient. This extension is reminiscent of the lopsided phenomenon seen in nearby \hi\ disks at $z\simeq0$ \citep{Baldwin1980, Sancisi2008}, which may be due to minor mergers and/or gas accretion events. Deeper CO observations are necessary to clarify the nature of this outer extension.

\begin{figure*}
\centering
\includegraphics[width=\textwidth]{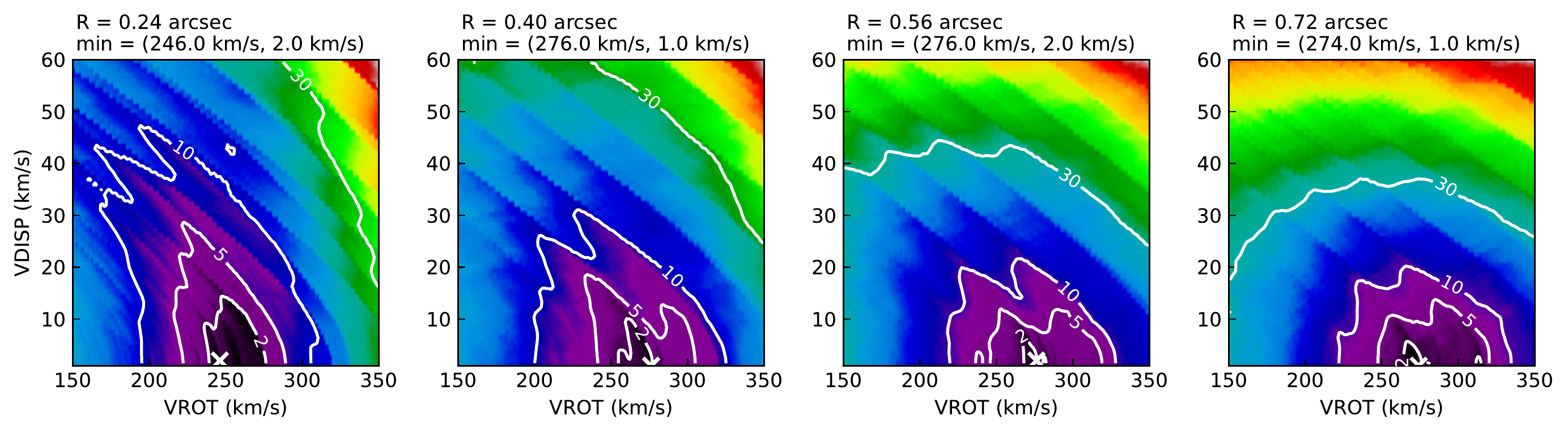}
\caption{The parameter space of rotation velocity (VROT) and velocity dispersion (VDISP) from modeling the H$\alpha$ cube of zC-400569 with \texttt{SpacePar}. The color-scale represents $\chi^2$ values at different radii, given in the top label. The best-fit values are indicated with a cross and reported in the top label. The white contours correspond to percentage variations in $\chi^2$ from the best-fit value (2\%, 5\%, 10\%, and 30\%). Similar plots are presented in Appendix\,\ref{app:3Dfits} for CO data. The small values of the velocity dispersion suggests that this quantity cannot be reliably measured because the line broadening is entirely dominated by the observational effects of spectral and spatial resolution.}
\label{fig:spacepar}
\end{figure*}

\section{Kinematic models}\label{sec:gaskin}

\subsection{Nonparametric 3D fitting}\label{sec:fitting}

We model the gas kinematics in a nonparametric fashion without a-priori assumption on the gas density profile, rotation curve shape, and gravitational potential. This is important because beam-smearing effects depend on several unknown quantities, such as the intrinsic gas distribution and rotation-curve shape \citep{Warner1973, Begeman1989}. Parametric 3D models assume predefined functions for the rotation-curve shape and the gas density profile, so they necessarily make implicit assumptions on the intrinsic strength of beam-smearing effects. For example, an intrinsically flat rotation curve down to small radii leads to more severe beam-smearing effects than an intrinsically rising rotation curve. As a result, parametric models may only partially account for beam-smearing effects. In addition, parametric models do not provide an actual derivation of the rotation curve because a smooth shape is imposed, neglecting possible real features (such as bumps and wiggles) that are often seen in rotation curves at $z\simeq0$ \citep[e.g.,][]{Sancisi2004, Lelli2016b}. On the other hand, parametric models have the advantage of having less free parameters than tilted-ring fits and may represent the only viable method to model rotating disks that are resolved with less than 2-3 resolution elements \citep[e.g.,][]{Bouche2015}.

The task \texttt{3Dfit} of \bb\ \citep{DiTeodoro2015} performs a tilted-ring modeling \citep{Warner1973, Rogstad1974, Begeman1989} fitting the 3D cube rather than 2D moment maps, so beam-smearing effects on each single line profile are taken into account. The disk is divided in a set of rings, where each ring is characterized by five geometric parameters $-$ center's coordinates ($x_0, y_0$), systemic velocity ($V_{\rm sys}$), position angle (PA), and inclination angle ($i$) $-$ and five physical parameters $-$ surface density ($\Sigma_{\rm gas}$), vertical thickness ($z_0$), rotation velocity ($V_{\rm rot}$), radial velocity ($V_{\rm rad}$), and velocity dispersion ($\sigma_{\rm V}$). 

We use rings with a width of $\sqrt{a\times b}/2$, where $a$ and $b$ are the major and minor axes of the synthesized beam. We adopt a fully axisymmetric disk, so the surface density of each ring is directly computed from the observed moment-zero map using azimuthal averages (option \texttt{Norm=Azim}). For the vertical density distribution, we assume an exponential law (option \texttt{Ltype=3}) with a fixed scale height of 100 pc: the precise value of $z_0$ is expected to have virtually no effects on our results considering the ALMA beam size of $3-4$ kpc. We also set $V_{\rm rad}=0$ because there is no strong evidence of radial motions in the velocity maps, such as a clear nonorthogonality between the kinematic major and minor axis (see, e.g., \citealt{Lelli2012a, Lelli2012b} for galaxies where $V_{\rm rad}\neq0$). Thus, we are left with seven free parameters in each ring. The systemic velocity is set to zero at the fiducial redshift of the galaxies (given in Table\,\ref{tab:cubes}) but is kept as a free parameter to account for minor adjustments in units of \kms.

For data with relatively low S/N and resolution (as for most high-$z$ galaxies), automated fitting codes such as \bb\ must be used with caution. Firstly, it is important to define a proper mask within which the residuals are evaluated. After various trials, we find that this mask should be conservative to avoid that pixels with low S/N bias the fit. To build such a mask, we use the task \texttt{Smooth \& Search} with parameters \texttt{Factor=1.5} (factor for spatial smoothing), \texttt{Snrcut=4} (primary S/N threshold), \texttt{Growthcut=3} (secondary S/N threshold when growing the initial mask), and \texttt{Minchannels=3} (minimum number of channels for a detection to be accepted, considering that the data have been Hanning smoothed over 3 channels). Next, our fitting strategy consists of three main steps: setting the disk geometry (Sect. \ref{sec:geo}), measuring the gas velocity dispersion (Sect.\,\ref{sec:disp}), and tracing the rotation curve (Sect.\,\ref{sec:rotcur}).

\subsection{Disk geometry}\label{sec:geo}

We run \texttt{3Dfit} on each cube leaving all seven parameters free. In this first fit, all pixels are uniformly weighted (option \texttt{Wfunc=0}). After various trials, we find that some best-fit parameters may slightly depend on some of the initial estimates, so we provide sensible initial guesses: $\mathrm{PA=35^{\circ}}$ and $i=50^{\circ}$ for zC-400569 and $\mathrm{PA=-20^{\circ}}$ and $V_{\rm sys}=50$ \kms\ for zC-488879. Initial estimates for all the other parameters are automatically estimated by \bb. Next, we determine the geometric parameters taking the median of the best-fit values of different CO lines across all rings; uncertainties are estimated as the median absolute deviation. Thus, we fix the same disk geometry for all emission lines using the CO kinematics. We do not model warps because there is no strong indication for them in any of the data cubes at the available resolution.

The \ci($1-0$)\ cube of zC-400569 has too low sensitivity to robustly determine the geometric parameters, while the H$\alpha$ kinematics may bias the results due to the H$\alpha$-bright companions to the south. In addition, the H$\alpha$ line could suffer from varying dust extinction across the disk, unlike the CO lines that trace the gas kinematics unbiased by such effects. The center and systemic velocity of the H$\alpha$ cube, however, are determined independently from CO lines to account for possible offsets in the astrometry and/or spectral calibration of the SINFONI and ALMA data. 

\begin{table}
\caption[]{Best-fit results from \bb. $V_{\rm sys}$ is measured with respect to the initial redshift estimate in Table\,\ref{tab:cubes}. $\langle V_{\rm rot} \rangle$ is the mean rotation speed; its uncertainty is estimated using Eq. 3 in \citet{Lelli2016a}.}
\centering
\footnotesize
\label{tab:3Dfits}
\renewcommand{\arraystretch}{1.3}
\begin{tabular}{lcc}
\hline
Parameter & zC-400569 & zC-488879 \\
\hline
$x_0$ (J2000)            & 150.28621$\pm$0.00002 & 150.28744$\pm$0.00001 \\
$y_0$ (J2000)            & 1.74116$\pm$0.00001   & 1.74118$\pm$0.00001\\
$V_{\rm sys}$ (\kms)    & 185$\pm$14 & 52$\pm$14 \\
PA ($^{\circ}$) & 37.5$\pm$8.6 & $-$19.2$\pm$6.3\\
$i$ ($^{\circ}$) & 54.5$\pm$5.0 & 71.6$\pm$4.1 \\
$\langle V_{\rm rot} \rangle$ (\kms)   & 254$\pm$41 & 336$\pm$29\\
$\sigma_{\rm CO}$ (\kms) & $\lesssim$15 & $\lesssim$15 \\
$\sigma_{\rm H\alpha}$ (\kms) & $\lesssim$37 & ... \\
\hline
\end{tabular}
\end{table}
Table\,\ref{tab:3Dfits} summarizes our results. The geometric parameters from CO kinematics are approximately consistent with the rest-frame optical morphology of the galaxies. The inclination of zC-400569, however, is more uncertain than the formal errors from the median absolute deviation: the CO($3-2$) line points to an inclined disk ($60-65^{\circ}$), while the CO($3-4$) and optical morphologies point to a more face-on disk ($45-50^{\circ}$). Thus, we assign an error of 5$^{\circ}$ that encompasses these inclinations within 2$\sigma$.

\begin{figure*}
\centering
\includegraphics[width=\textwidth]{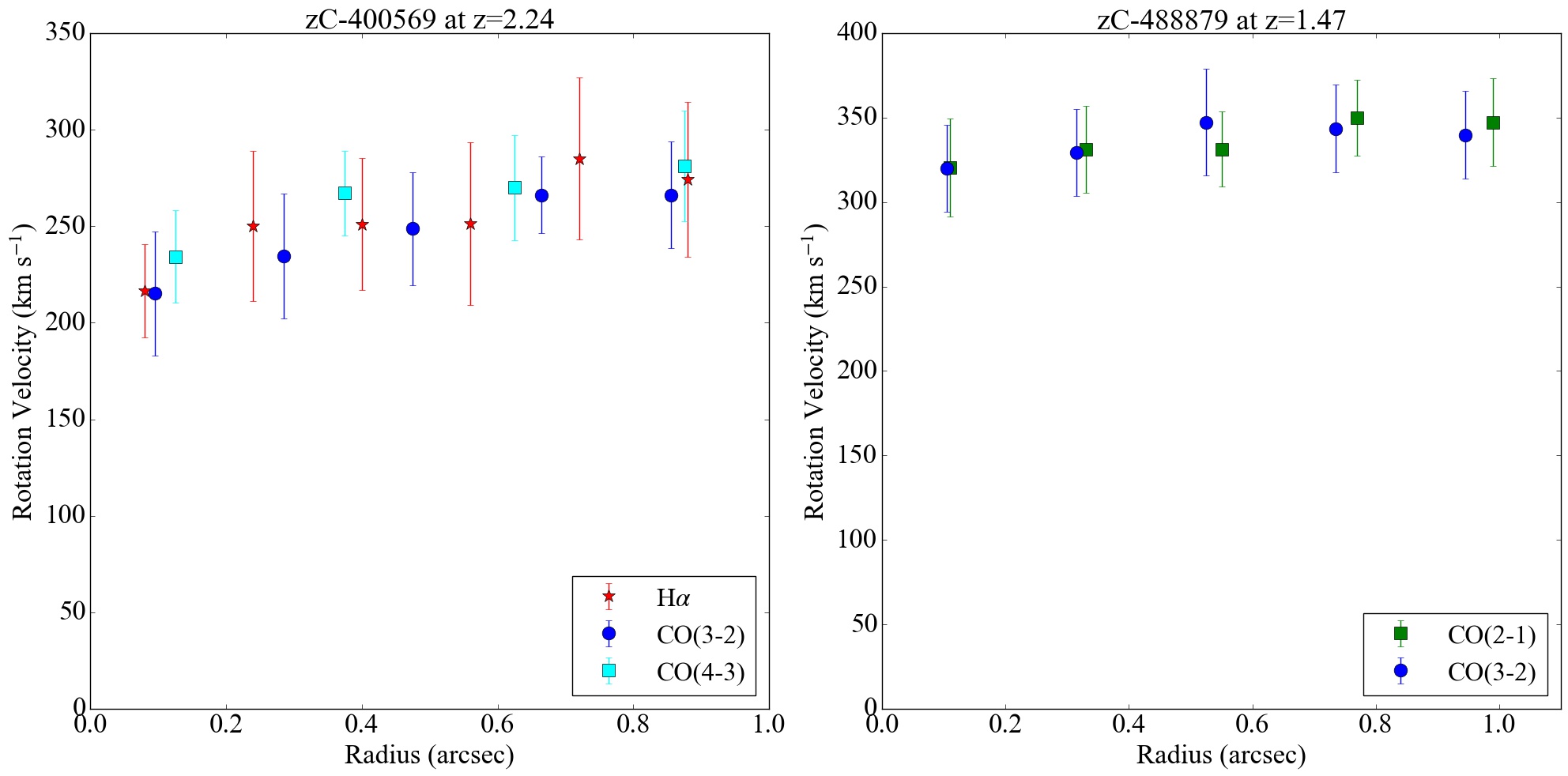}
\caption{Rotation curves of zC-400569 (left) and zC-488879 (right). Different symbols show velocity measurements from different emission lines: green squares for CO($2-1$), blue circles for CO($3-2$), cyan squares for CO($4-3$), and red stars for H$\alpha$.}
\label{fig:rotcur}
\end{figure*}

\subsection{Gas velocity dispersion}\label{sec:disp}

Having fixed the disk geometry, we use the task \texttt{SpacePar} in \bb\ to look for a global minimum in the $V_{\rm rot}-\sigma_{\rm V}$ space. The parameter space is explored in steps of 1 \kms, considering a range in $\sigma_{\rm V}$ from 1 to 60 \kms\ and a range in $V_{\rm rot}$ of 200 \kms\ centered on the expected rotation velocity. At each step, \texttt{SpacePar} builds a 3D disk model with given $V_{\rm rot}$ and $\sigma_{\rm V}$, then computes the average sums of residuals $\Delta r = |M - D|$, where $M$ and $D$ are the flux values at each 3D pixel of model and data, respectively (default option \texttt{Ftype=2}). All pixels are uniformly weighted (option \texttt{Wfunc=0}).

For all emission lines, {\texttt{SpacePar} returns} very low values of $\sigma_{\rm V}$ of just $1-2$ \kms. For example, Fig.\,\ref{fig:spacepar} shows the results for the H$\alpha$ cube of zC-400569, excluding the innermost ring in which beam smearing effects are most severe and the outermost ring in which the S/N is low. Similar plots for the CO cubes are shown in Appendix\,\ref{app:3Dfits}. Values of $\sigma_{\rm V}\simeq1-2$ \kms\ are physically acceptable but the minimum in the $V_{\rm rot}-\sigma_{\rm V}$ space is very shallow; an intrinsic velocity dispersion of $\sim$10 \kms\ increases the $\chi^2$ by a mere 5$\%$. Most likely, the available spectral resolution ($\Delta_{\rm V}$), spatial resolution, and sensitivity do not allow us to robustly measure the intrinsic velocity dispersion.

One may wonder whether the low velocity dispersions are driven by high surface brighntess pixels, which could dominate the average sum of residuals $\Delta r$. We reran \texttt{SpacePar} calculating $\Delta {\hat{r}} = |M - D|/|M+D|$, which normalizes the residuals at each pixel to the local flux density (option \texttt{Ftype=3}). Again, we found formal best-fit values of $\sigma_{\rm V}$ of just 1-2 \kms, indicating that this result is not driven by how we weight the residuals.

To further explore the situation and set a fiducial upper limit on $\sigma_{\rm V}$, we built mock-observed cubes of rotating disks with known $V_{\rm rot}$ and $\sigma_{\rm V}$. We smoothed the mock cubes to the same spatial and spectral resolutions of our observations, and add Gaussian noise with a peak S/N of 6 to 8, similarly to our data. We then ran \texttt{SpacePar} on each mock-observed cube. Independently of the input value of $V_{\rm rot}$, when the intrinsic $\sigma_{\rm V}$ is equal or smaller than the instrumental dispersion $\sigma_{\rm inst}\simeq \Delta_{\rm V}/2.35$, \texttt{SpacePar} pushes the best-fit value of $\sigma_{\rm V}$ to the lower boundary of the explored parameter space. This is the same behavior that we obtain for all emission-line cubes of zC-400569 and zC-488879. When the intrinsic $\sigma_{\rm V}$ is slightly higher than $\sigma_{\rm inst}$, instead, \texttt{SpacePar} does \textit{not} push $\sigma_{\rm V}$ to the lower boundary of the explored parameter space, albeit it may not exactly recover the correct $\sigma_{\rm V}$ due to the low S/N. Thus, we conclude that $\sigma_{\rm inst}$ is a sensible upper limit for the intrinsic velocity dispersion.

Given that the ALMA cubes have a spectral resolution $\Delta_{\rm V} \simeq 30$ \kms, the formal upper limit on the CO velocity dispersion ($\sigma_{\rm CO}$) is about 13 \kms. To be conservative, we adopt the channel width of 15 \kms\ as our upper limit on $\sigma_{\rm CO}$. Given the SINFONI spectral resolution $\Delta_{\rm V} \simeq86$ \kms\ around the H$\alpha$ line \citep{Forster2009}, we adopt an upper limit of 37 \kms on the H$\alpha$ velocity dispersion ($\sigma_{\rm H\alpha}$). We stress that these upper limits are themselves uncertain: while the low spectral and spatial resolutions increase the line broadening leading to an overestimate of the intrinsic $\sigma_{\rm V}$, the low S/N may lead to an underestimate of the intrinsic $\sigma_{\rm V}$.

To check the robustness of our upper limits, we created observed cubes with a channel width of 5 \kms\ and spectral resolution of 10 \kms\ (after Hanning smoothing). These cubes have significantly lower sensitivity than the cubes at 30 \kms\ resolution. In fact \bb\ does not detect the source in most of them apart from the CO($4-3$) cube of zC-400569 and the CO($3-2$) cube of zC-488879. For these two CO cubes, \texttt{SpacePar} provides $\sigma_{\rm CO}\simeq4-12$ \kms, which may be considered as proper measurements because $\sigma_{\rm inst}\simeq 4$ \kms. Given the low S/N of these cubes, however, we prefer to be conservative and base all our conclusions on the upper limit $\sigma_{\rm CO}<15$ \kms.

\subsection{Rotation curves}\label{sec:rotcur}

Fixing the velocity dispersion to its formal upper limit, we run \texttt{3Dfit} using only $V_{\rm rot}$ as a free parameter in each ring. This provides the observed rotation curve for each emission line. In this last fit, pixels are weighted according to the function $cos^2(\theta)$, where $\theta$ is the azimuthal angle around the major axis (option \texttt{Wfunc=2}). Thus, pixels around the major axis are maximally weighted to maximize information on rotational velocities.

The right panels of Fig.\,\ref{fig:Mosaic1} and Fig.\,\ref{fig:Mosaic2} compare major-axis PV diagrams from the observed cubes with those from our best-fit 3D disk models. The best-fit rotation curves are projected on the PV diagrams as $V_{\rm rot}\sin(i)$. Overall, the 3D disk models give a good description of the observations. In particular, the observed thickness of the PV diagrams (the line broadening at each radius) is well reproduced, indicating that the assumed velocity dispersion is sensible. The bulk of the observed line broadening, indeed, is driven by beam-smearing effects, not by the intrinsic gas velocity dispersion. Another known effect of severe beam smearing is that the best-fit rotation velocities are not necessarily near the peak of the emission line along the major axis, but they are closer to the uppermost and lowermost velocity edges of the PV diagrams. Similar rotation curves, indeed, could be obtained using the ``envelope-tracing method'', which considers the terminal velocity near the lowest density contour in a PV diagram \citep[e.g.,][]{Sofue2001}. The envelope-tracing method was originally developed for edge-on galaxies \citep{Sancisi1979} but can be applied in poorly resolved galaxies too because multiple line profiles are flux-averaged within the PSF (similarly to the line-of-sight integration in edge-on disks), so the true rotation velocity corresponds to the terminal velocity near the edge of the emission, not the velocity near the peak of the emission.

The rotation curves are shown in Figure\,\ref{fig:rotcur}. Reassuringly, different emission lines provide consistent results, suggesting that the various gas phases are kinematically settled. The observed rotation velocities imply $V_{\rm rot}/\sigma_{\rm CO}\gtrsim17$ for zC-400569 and $V_{\rm rot}/\sigma_{\rm CO}\gtrsim22$ for zC-488879, so there is no need to apply corrections for pressure support (asymmetric drift). Overall, the different gas tracers point to the same result: flat rotation curves with no clear signs of a decline in the outer parts.

\bb\ provides asymmetric errors ($\delta_{+}$, $\delta_{-}$) that correspond to a variation of 5$\%$ of the residuals from the global minimum. We compute symmetric 1$\sigma$ errorbars as $\sqrt{\delta_{+}\delta_{-}}$, which will be used as weights in subsequent Bayesian fits (Sect.\,\ref{sec:priors}). For zC-400569, rotation velocities from the CO($4-3$) line have the smallest uncertainties, while those from the H$\alpha$ have the largest ones due to the relative S/N of the data. For zC-488879, rotation velocities from the CO($2-1$) and CO($3-2$) lines have comparable uncertainties. 

\subsection{Comparison with previous H$\alpha$ studies}

The H$\alpha$ kinematics of zC-400569 has been previously studied by the SINS/zC-SINF team using the \textsc{Dysmal} parametric code \citep{Genzel2017, Genzel2020, Price2021, Nestor2022}. The previous H$\alpha$ results are mixed in terms of both the intrinsic H$\alpha$ velocity dispersion and the rotation curve shape.

Regarding the H$\alpha$ velocity dispersion, we give a formal upper limit of 37 \kms\ while the SINS/zC-SINF team reports actual measurements: 34 \kms \citep[][consistent with our upper limit]{Genzel2017}, 45 \kms\, \citep{Genzel2020}, $71.82_{-8.40}^{+5.97}$ \kms\, \citep{Price2021}, and $58\pm5$ \kms \citep{Nestor2022}. We do not fully understand why the SINS/zC-SINF team finds significantly different velocity dispersions from the same IFU data, but these varying values are in line with our basic conclusion: the H$\alpha$ velocity dispersion of zC-400569 cannot be reliably measured with the existing data. This demonstrates that fitting methodologies can play a major role in obtaining measurements and/or upper limits of $\sigma_{\rm V}$ from poorly resolved data with low S/N.

Regarding the H$\alpha$ rotation curve, our measurements and previous ones differ in both the inner and outer parts. In the inner parts, the H$\alpha$ rotation curve from \textsc{Dysmal} fits \citep{Genzel2017, Genzel2020, Price2021, Nestor2022} rises until $R\simeq0.4''$ whereas our rotation curve is already flat. As far as we understand, the previous H$\alpha$ works show the beam-smeared rotation curve, not the intrinsic one such as that provided by \bb\ (Fig.\,\ref{fig:rotcur}), so the differences in the inner rotation velocities can be ascribed to beam-smearing effects \citep[e.g.,][]{Lelli2010}. In the outer parts, the H$\alpha$ rotation curve from \textsc{Dysmal} 1D fits \citep{Genzel2017, Genzel2020, Nestor2022} shows a mild decline, whereas that from \textsc{Dysmal} 2D fits \citep{Price2021} keeps rising until the last measured point at $R\simeq1''$. Our 3D fits do not confirm any of these trends, but give a flat rotation curve out to $R\simeq1''$. These differences are likely driven by the H$\alpha$-bright companions to the south (Fig.\,\ref{fig:SINFONI}) that may hinder a precise determination of the PA, as also pointed out by \citet{Price2021}. The CO data do not suffer from these complications and give a PA of 37.5$^{\circ}$; the resulting major-axis PV diagrams (Figure \ref{fig:Mosaic1}) unambiguously show that the rotation curve is relatively flat. In fact, independently of fitting codes, one would obtain the same result by tracing the rotation curve by eye using a ``human neural network'' \citep[e.g.,][]{Sancisi1979, Verheijen2001}.

The different measurements of $\sigma_{\rm V}$ and $V_{\rm rot}$ from different studies imply different values of the ratio $V_{\rm rot}/\sigma_{\rm V}$, which determine whether pressure support is relevant or not. Differently from previous studies, we conclude that the values of $V_{\rm rot}/\sigma_{\rm V}$ are relatively high ($\gtrsim$7 for H$\alpha$ and $\gtrsim17$ for CO), so corrections for pressure support (asymmetric drift) are negligible. Thus, the observed rotation speeds trace the circular velocity of a test particle in the equilibrium gravitational potential and can be safely used to build mass models. In the case of zC-400569, the value of $\sigma_{\rm V}$ plays the key role because the absolute differences in $V_{\rm rot}$ from different studies are relatively small, apart from the overall difference in rotation-curve shape.

\begin{table*}
\caption[]{Results from Bayesian fits to the rotation curves for different mass models. For the uncertainties, see Appendix\,\ref{app:corner} because several posterior-probability distributions are not Gaussian, so formal $\pm1\sigma$ errors can be misleading.}
\centering
\footnotesize
\label{tab:mass}
\renewcommand{\arraystretch}{1.3}
\begin{tabular}{ccccccccccc}
\hline
\noalign{\smallskip}
Galaxy & Model & $i$ & $M_{\rm gas}$ & $M_{\rm disk}$ & $M_{\rm bul}$ & $M_{200}$ & $C_{200}$ & $M_{\rm bar}$ & $M_{\rm bul}/M_{\rm bar}$\\
&      & (deg.) & \multicolumn{3}{c}{---------- (10$^{10}$ M$_\odot$) ----------} & (10$^{12}$ M$_\odot$) & & (10$^{10}$ M$_\odot$) & \\
\hline
zC-400569 & Baryons only & $54.7_{-5.0}^{+5.0}$ & $4.9_{-1.6}^{+1.3}$ & $5.3_{-2.8}^{2.3}$ & $2.5_{-0.5}^{+0.5}$ & ... & ... & $12.8_{-3.3}^{+2.7}$ & 0.20\\
& Baryons$+$NFW & $54.6_{-4.8}^{+5.0}$ & $2.3_{-1.4}^{+1.0}$ & $2.2_{-1.5}^{+1.2}$ & $2.4_{-0.5}^{+0.5}$ & $2.2_{-0.7}^{+0.8}$ & $4.2_{-1.4}^{+1.4}$ &
$6.9_{-2.1}^{+1.7}$ & 0.35\\
& MOND & $54.8_{-5.0}^{+4.9}$ & $3.7_{-1.3}^{+1.1}$ & $3.8_{-2.1}^{+1.7}$ & $2.6_{-0.5}^{+0.5}$ & ... & ... & $10.2_{-2.5}^{+2.1}$& 0.25\\
zC-488879 & Baryons only & $71.6_{-4.0}^{+4.1}$ & $5.6_{-1.7}^{+1.4}$ & $6.6_{-3.9}^{+3.3}$ & $14.6_{-1.6}^{+1.5}$ & ... & ... & $26.8_{-4.5}^{+3.9}$ & 0.54\\
& Baryons$+$NFW & $71.5_{-4.0}^{+4.1}$ & $1.8_{-1.4}^{+1.0}$ & $1.7_{-1.6}^{+1.2}$ & $13.4_{-1.5}^{+1.4}$ & $15.3_{-6.5}^{+7.2}$ & $3.7_{-1.2}^{+1.1}$ & $16.9_{-2.6}^{+2.1}$ & 0.79\\
& MOND & $71.8_{-4.0}^{+4.1}$ & $3.9_{-1.4}^{+1.2}$ & $4.3_{-2.8}^{+2.4}$ & $14.6_{-1.5}^{+1.5}$ & ... & ... & $22.8_{-3.5}^{+3.1}$ & 0.64\\
\hline
\end{tabular}
\end{table*}

\section{Mass models}\label{sec:massmodels}

\subsection{Baryonic gravitational contributions}\label{sec:vbar}

We fit the rotation curves with mass models that includes different gravitational contributions. We start with mass models that contain only baryonic components (Sect.\,\ref{sec:onlybaryons}), then mass models that also include a dark matter (DM) halo (Sect.\,\ref{sec:darkmatter}), and finally mass models in the context of Milgromian dynamics (MOND; Sect.\,\ref{sec:MOND}). In this section we describe the calculation of the Newtonian baryonic contributions: $V_{\rm bul}$ from the stellar bulge, $V_{\rm disk}$ from the stellar disk, and $V_{\rm gas}$ from the cold gas disk.

The contribution of each baryonic component is computed using the task \textsc{Rotmod} in the \texttt{Gipsy} software \citep{Vogelaar2001}. The total baryonic contribution is then given by
\begin{equation}\label{eq:Vmod}
 V_{\rm bar}^2 = \sum_{i} \Upsilon_{\rm i}V_{\rm i}^2 \quad \rm{with}\quad i=\rm{bul,\,disk,\,gas}
\end{equation}
where $\Upsilon_{\rm i} = M_{\rm i}/(10^{10} M_{\odot})$ are dimensionless factors. Basically, for numerical convenience, the gravitational contribution of each component is computed for an arbitrary mass of 10$^{10}$ M$_\odot$ and rescaled using fitting parameters $\Upsilon_{\rm i}$ on the order of one. The total baryonic mass of the model is then given by
\begin{equation}
M_{\rm bar} = (\Upsilon_{\rm bul} + \Upsilon_{\rm disk} + \Upsilon_{\rm gas}) \times 10^{10} \, \rm{M}_\odot.
\end{equation}

The gravitational contribution of the stellar bulge is computed assuming spherical symmetry, deprojecting a given 2D surface density profile \citep{Kent1986}. We adopt a \citet{DeVauc1948} profile for which $R_{\rm e,\,bul}$ is fixed from fits of the observed surface brightness profile (Sect.\,\ref{sec:HST}). Thus, the only free parameter is $M_{\rm bul}$ or equivalently $\Upsilon_{\rm bul}$. Importantly, it is necessary to extrapolate the bulge profile at radii smaller than the HST spatial resolution ($\sim$1.3 kpc) otherwise the innermost points of the rotation curve would not be correctly reproduced. Differences in $V_{\rm bul}$ between spherical and oblate geometries are on the order of $10\%-20\%$ \citep{Noordermeer2008} and degenerated with $M_{\rm bul}$, which is much more uncertain than that.

The gravitational contribution of the stellar disk is computed considering a disk of finite thickness with a given density profile $\rho(R, z) = \Sigma(R)\zeta(z)$ \citep{Casertano1983}. For the radial density distribution $\Sigma(R)$, we use the best-fit model to the observed surface brightness profile after subtracting the bulge component (Sect.\,\ref{sec:HST}). Thus, for zC-400569 we adopt a pure exponential disk, while for zC-488879 the disk term includes an exponential plus a barlens-like component (see Fig.\,\ref{fig:SBprof}). Differently from mass modeling at $z\simeq0$ \citep{Lelli2016b}, we use parameteric models for the radial density profile because they need to be extrapolated at $R<1.3$ kpc due to limited spatial resolution of the available images. For the vertical density distribution $\zeta(z)$, we assume an exponential function with constant scale height $z_{\rm disk}$. We estimate $z_{\rm disk}$ using the scaling relation $z_{\rm disk} = 0.196 \, (R_{\rm e, disk}/1.68)^{0.633}$ that holds for edge-on disk galaxies at $z\simeq0$ \citep{Bershady2010}. The vertical geometry has a small effect on the resulting $V_{\rm disk}$: at fixed $M_{\rm disk}$ a thick disk gives a smaller velocity contribution than a thin disk at small radii \citep{Casertano1983}. The uncertainties on $M_{\rm disk}$, however, are significantly larger than plausible variations in $z_{\rm disk}$, so there is little value in leaving $z_{\rm disk}$ free in the fit.

The gravitational contribution of the gas disk is computed in a similar fashion as the stellar disk. For the radial density distribution, we use the observed CO($3-2$) profile for zC-400569 and the observed CO($2-1$) profile for zC-488879 because they are the lowest CO transitions available, which are expected to best trace the H$_2$ surface density distribution irrespective of the ambient conditions (such as gas density and kinetic temperature). For the vertical density distribution, we assume an exponential profile with a constant scale height of 100 pc, which is reasonable for a cold gas disk with $V_{\rm rot}/\sigma_{V} \gtrsim 20$. Our computation of $V_{\rm gas}$ neglects the contribution of atomic gas (\hi), which may dominate the total gas budget of high-$z$ galaxies \citep{Chowdhury2022}. In the inner regions of local spirals, however, the H$_2$ surface densities usually dominate over the \hi\ surface densities, so the gas gravitational contribution is mostly due to H$_2$ at $R\lesssim5-10$ kpc while \hi\ prevails at larger radii \citep{Martinsson2013, Frank2016}. According to theoretical models of the interstellar medium (ISM), the transition from atomic to molecular gas depends mostly on local properties (such as pressure, metallicity, and far-UV radiation) and is expected to be spatially abrupt once the \hi-to-H$_2$ phase transition criterion is satisfied, leading to a nearly fully molecular ISM \citep{Elmegreen1989, Elmegreen1993, Papadopoulos2002, Offner2013}. Given that we extract rotation curves using CO lines, it is safe to assume that these CO-bright regions are well within the \hi-to-H$_2$ phase transition radius, thus dominated by molecular gas \citep{Bisbas2021}.

\subsection{Bayesian likelihood and priors}\label{sec:priors}

The parameters of the mass models are determined using a Markov-Chain-Monte-Carlo (MCMC) method in a Bayesian context. We define the likelihood $\mathcal{L} = \exp({-0.5 \chi^2})$ with
\begin{equation}\label{eq:chi}
 \chi^2 = \sum_{k}^{N} \frac{[V_{\rm rot} - V_{\rm mod}(\vec{p})]^{2}}{\delta^{2}_{V_{\rm rot}}},
\end{equation}
where $V_{\rm rot}$ is the observed rotation velocity at radius $R_k$, $\delta_{V_{\rm rot}}$ is the associated error, and $V_{\rm mod}$ is the model rotation velocity that depends on the fitting parameters $\vec{p}$. As it is often the case in Astronomy, the errors $\delta_{V_{\rm rot}}$ are ``educated guesses'' and do not truly represent a strict 1$\sigma$ deviation from a Gaussian distribution. However, scaling all $\delta_{V_{\rm rot}}$ by an arbitrary factor would only affect the width of the posterior probability distribution but not its overall shape. This implies that we can robustly determine the best-fit model parameters $\vec{p}$, but the associated uncertainties should be taken with a grain of salt. For each galaxy, we consider the ensemble of rotation velocity measurements from different emission lines at the same time (Fig.\,\ref{fig:rotcur}). This increases the statistical significance of our data set, allowing us to more robustly infer the optimal model parameters and associated uncertainties.

The posterior probability distributions of the model parameters are mapped using \texttt{emcee} \citep{Foreman2013}. The MCMC chains are initialized with 200 walkers. We run 1000 burn-in iterations, then the sampler is run for another 2000 iterations. The \texttt{emcee} parameter $a$, which controls the size of the stretch move, is set equal to 2. In general, this gives acceptance fractions larger than 50\%.

Both $V_{\rm rot}$ and $\delta_{V_{\rm rot}}$ depend on the disk inclination $i$. The value of $i$ is constrained by kinematic fits to the emission-line cubes with uncertainties of a few degrees (Sect.\,\ref{sec:geo}). Thus, we treat $i$ as a nuisance parameter using a Gaussian prior with central value $i_0$ and standard deviation $\delta_{i_0}$ set equal to the kinematic estimates from \bb. Then, $V_{\rm rot}$ and $\delta_{V_{\rm rot}}$ transform as $\sin(i_0)/\sin(i)$.

Fitting rotation curves can be a strongly degenerate problem \citep[e.g.,][]{Li2019, Li2020, Li2021}. For both galaxies, indeed, the velocity contributions $V_{\rm gas}$ and $V_{\rm disk}$ display a similar trend with radius, so the values of $\Upsilon_{\rm gas}$ and $\Upsilon_{\rm disk}$ are strongly degenerate and essentially unconstrained when left entirely free. To break this degeneracy, we impose two physically motivated priors:
\begin{enumerate}
    \item A log-normal prior on $\Upsilon_{\rm gas}$ using order-of-magnitude estimates of $M_{\rm gas}$ from CO luminosities. Adopting the average line ratios $R_{21}=L_{\rm CO(2-1)}/L_{\rm CO(1-0)}\simeq0.85$ and $R_{31}=L_{\rm CO(3-2)}/L_{\rm CO (1-0)}\simeq0.60$ \citep[e.g.,][]{Bisbas2021}, we find $M_{\rm gas}\simeq10^{10}-10^{11}$ M$_\odot$ assuming either a starburst-like conversion factor $\alpha_{\rm CO}\simeq0.4$ M$_\odot$ (K km s$^{-1}$ pc$^2$)$^{-1}$ or the Milky Way $\alpha_{\rm CO}\simeq4.3$ M$_\odot$ (K km s$^{-1}$ pc$^2$)$^{-1}$. Thus, we center the prior at $\log(\Upsilon_{\rm gas})=0.5$ with a standard deviation of 0.5 dex.
    \item A log-normal prior on $\mu_{\rm gas}=\Upsilon_{\rm gas}/\Upsilon_{\rm disk}$, the ratio between gas mass and stellar mass in the star-forming disk. Considering the results of \citet{Tacconi2018} and \citet{Liu2019b} for galaxies at $z\simeq1.5-2.5$, we center the prior at $\log(\mu_{\rm gas})=0$ with a standard deviation of 0.3 dex.
\end{enumerate}
These priors are used in all the mass models described in the following sections. The fitting results are summarized in Table\,\ref{tab:mass}.

\begin{figure*}
\centering
\includegraphics[width=0.33\textwidth]{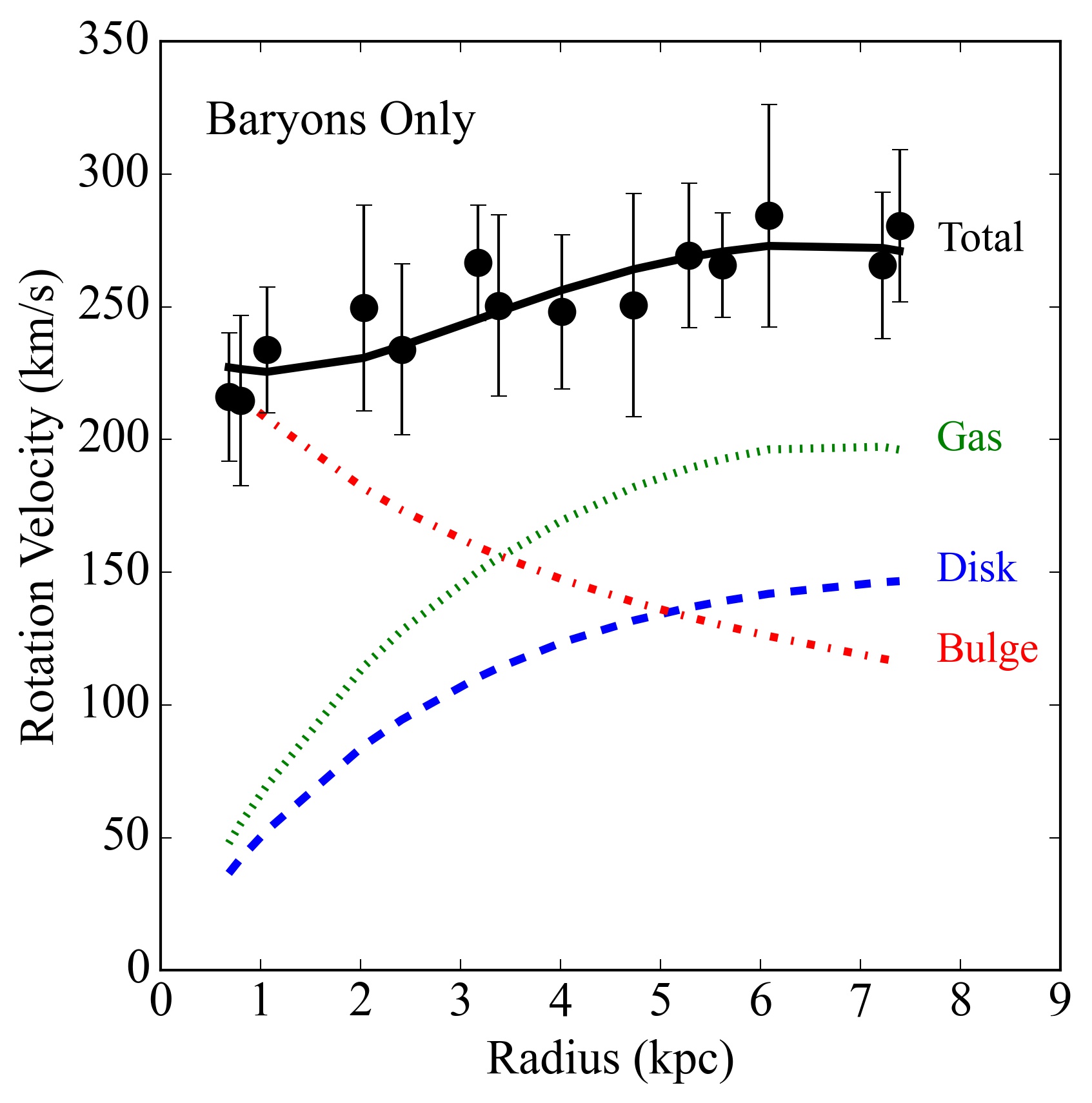}
\includegraphics[width=0.33\textwidth]{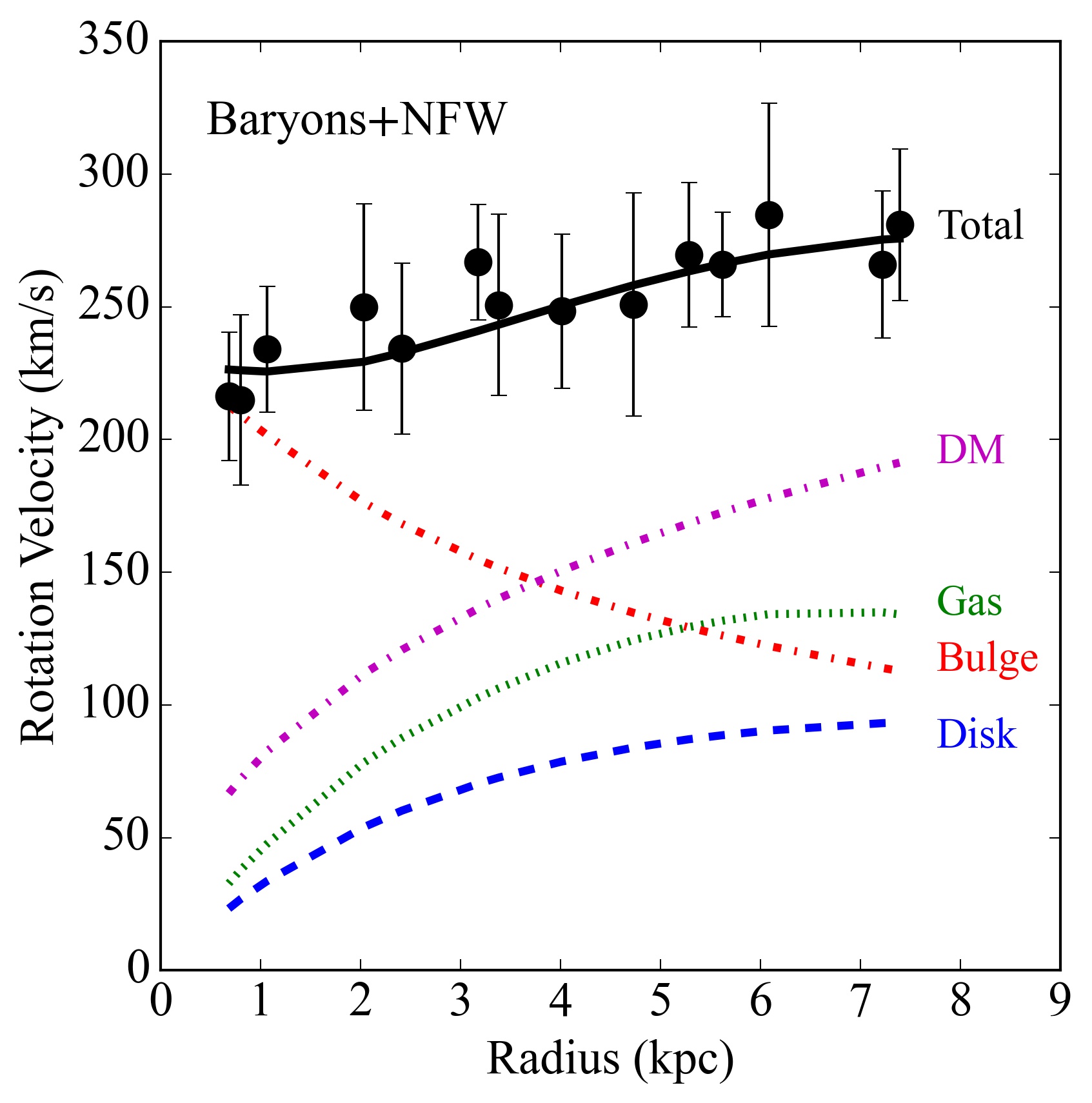}
\includegraphics[width=0.33\textwidth]{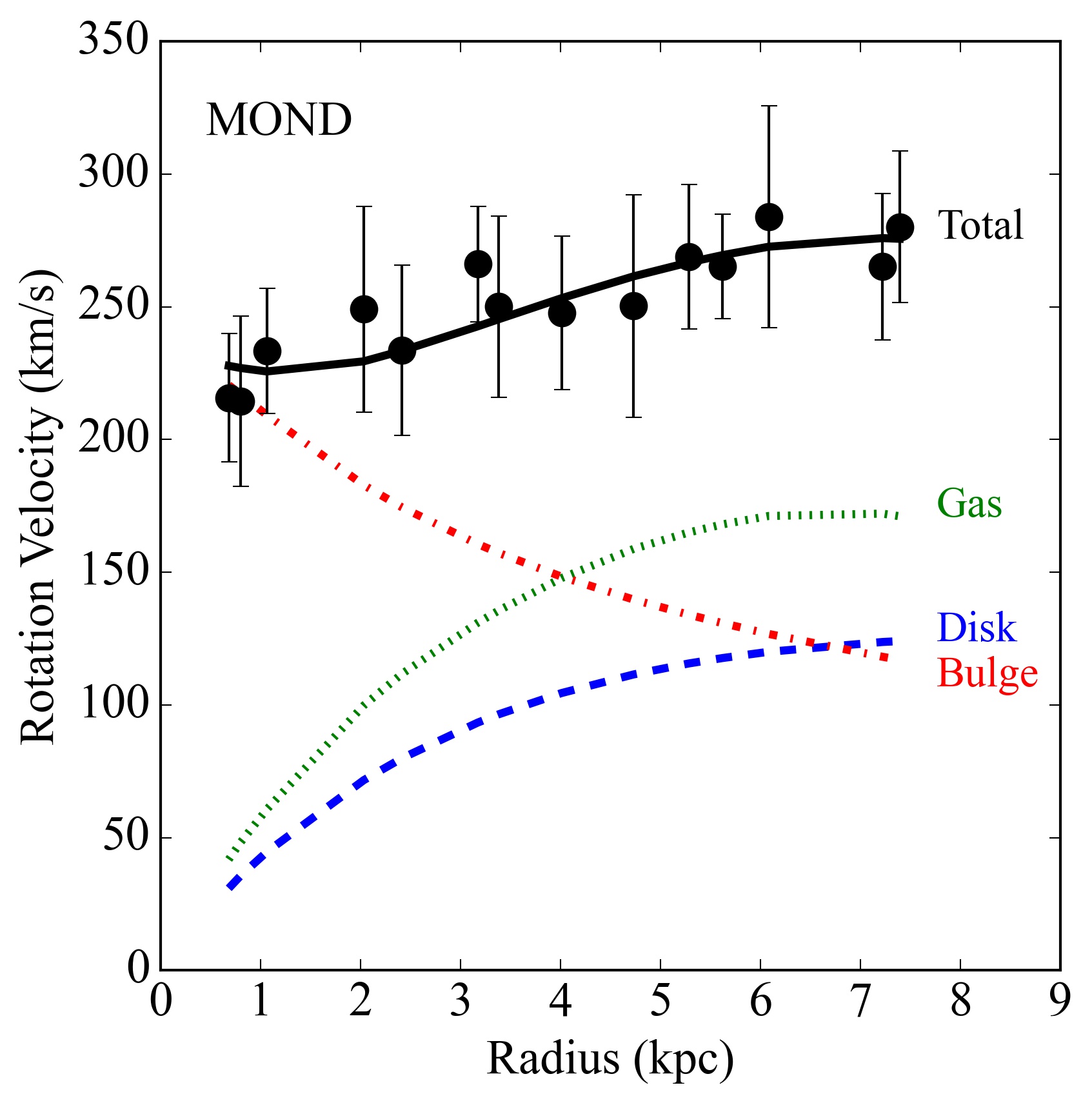}
\includegraphics[width=0.33\textwidth]{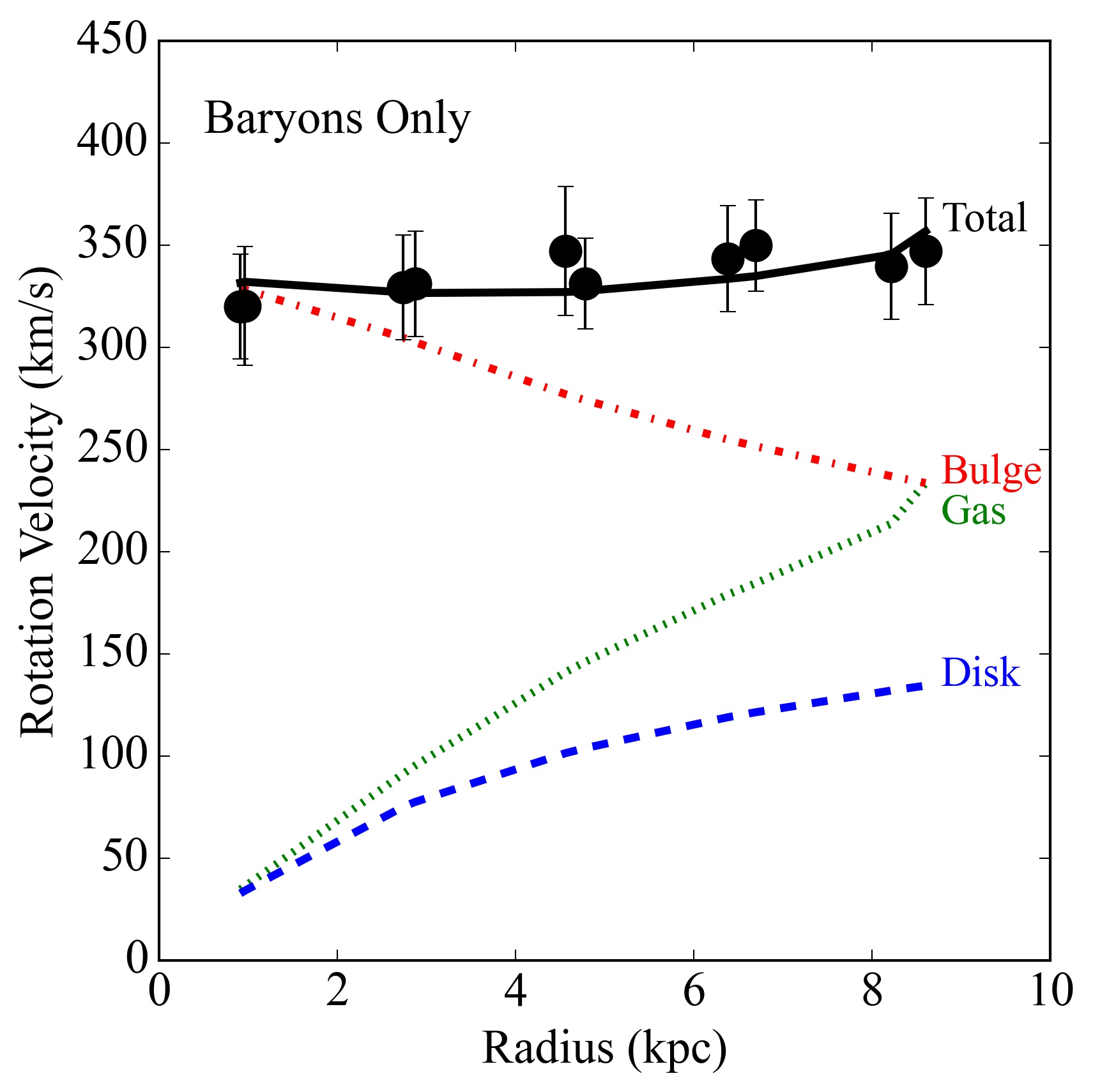}
\includegraphics[width=0.33\textwidth]{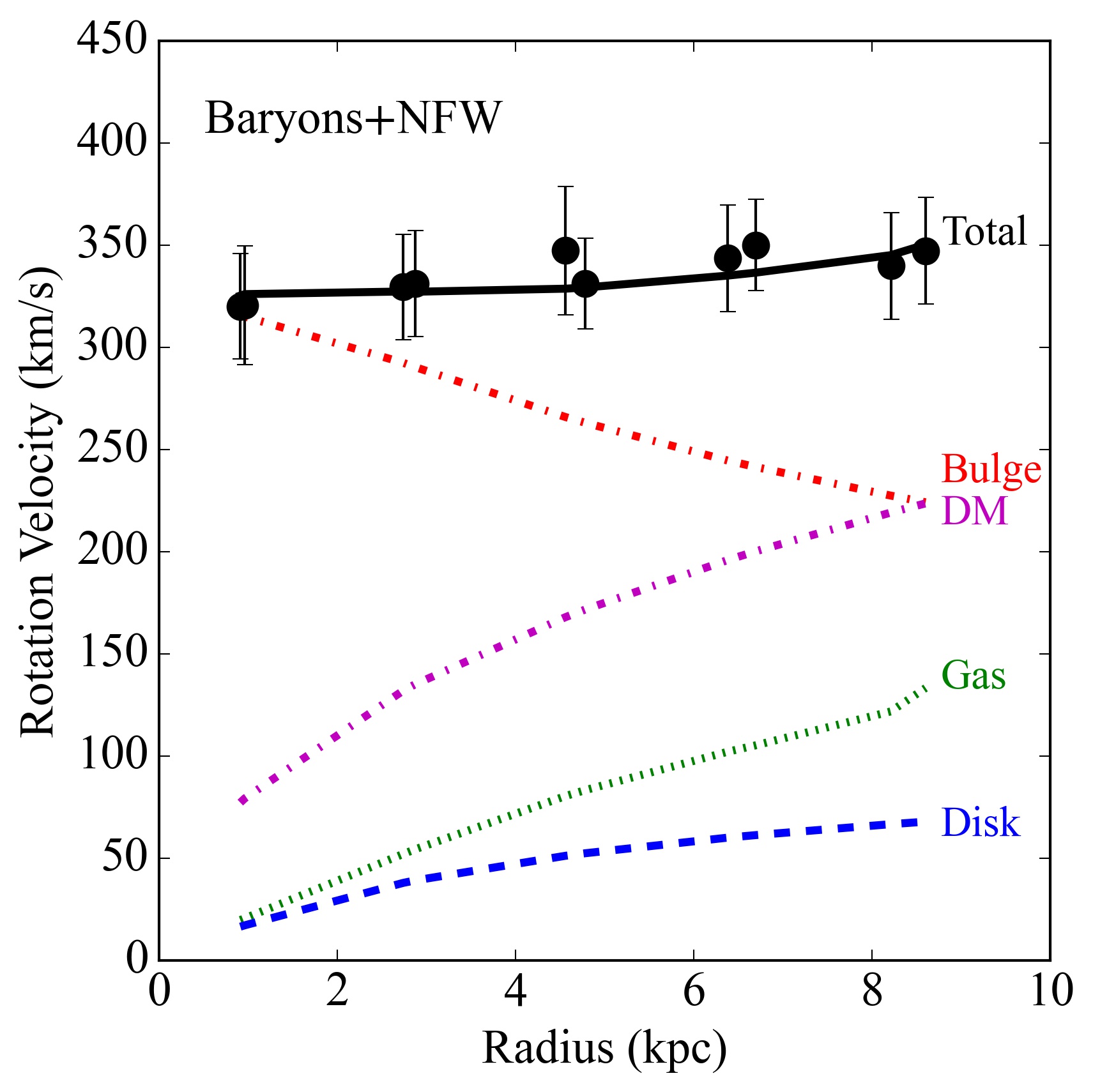}
\includegraphics[width=0.33\textwidth]{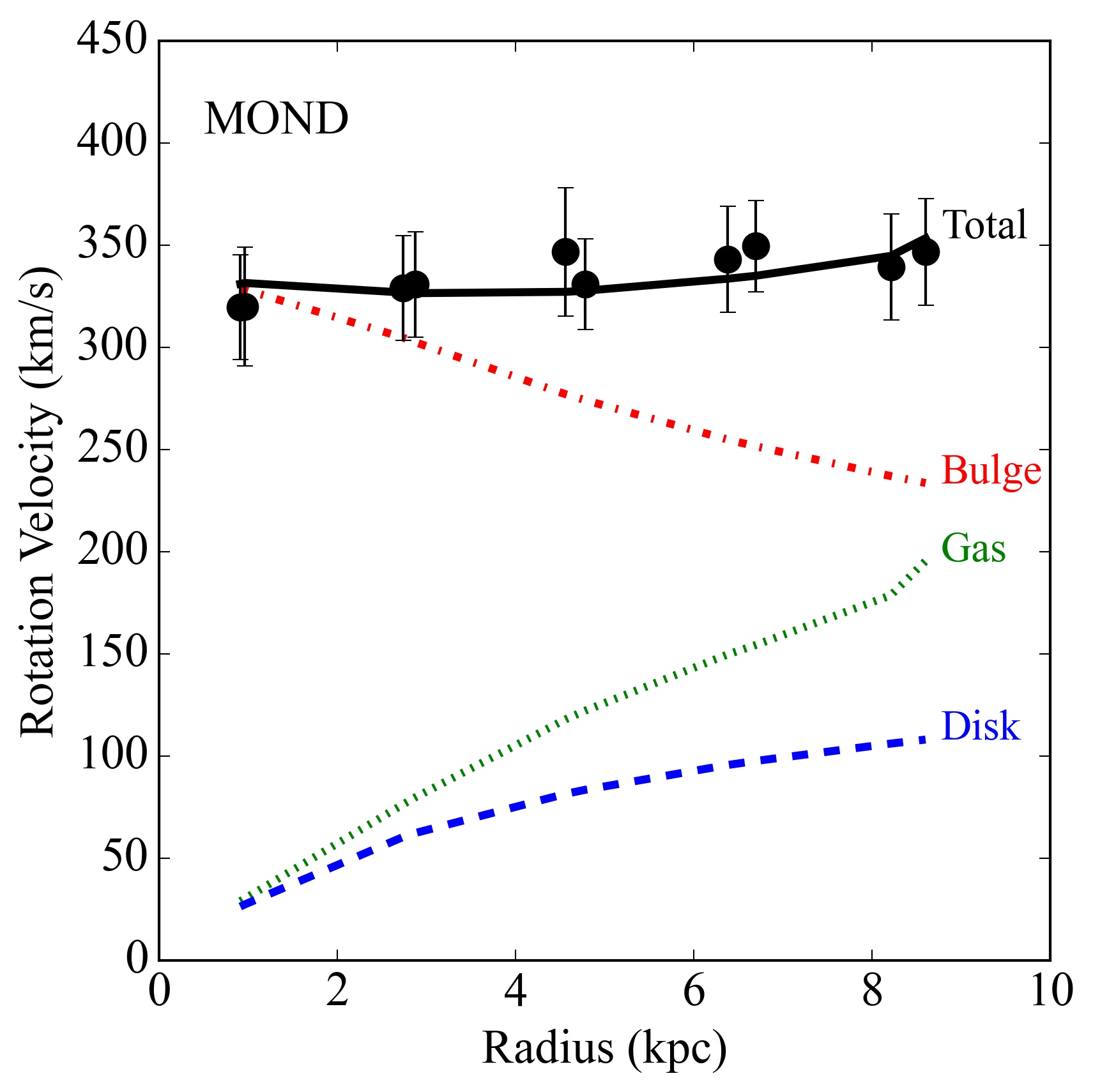}
\caption{Mass models for zC-400569 (top panels) and zC-488879 (bottom panels) using only baryonic components (left), baryons plus a NFW halo (middle) and baryons in a MOND context (right). In all panels, the observed rotation curve (dots with errorbars) is fitted considering the gravitational contributions of a gas disk (green dotted line), a stellar disk (blue dashed line), a stellar bulge (red dash-dotted line, if present), and a DM halo (magenta dash-dotted line, if present).}
\label{fig:MassModels}
\end{figure*}

\subsection{Mass models with baryons only}\label{sec:onlybaryons}

Figure\,\ref{fig:MassModels} (left panels) shows mass models considering only baryonic components, so $V_{\rm mod} = V_{\rm bar}(\Upsilon_{\rm bul}, \Upsilon_{\rm disk}, \Upsilon_{\rm gas})$ in Eq.\,\ref{eq:chi}. Given the overwhelming evidence for the DM effect in the Universe, these mass models are not entirely physical but provide hard upper limits on stellar and gas masses; they are analogous to ``maximum disk'' mass models of local galaxies \citep{vanAlbada1985, Starkman2018}. For both galaxies, the rotation curve is well reproduced with no need of DM. Similarly to local spiral galaxies, the bulge dominates the inner galaxy regions, while the stellar and gas disks become important at larger radii. Differently from local galaxies, however, the molecular gas disk dominates over the stellar disk. This result is partially driven by our assumed priors  (see Sect.\,\ref{sec:priors}): the relative contributions of $V_{\rm gas}$ and $V_{\rm disk}$ would be nearly unconstrained if left entirely free. The ratio $M_{\rm bul}/M_{\rm bar}$, instead, is a robust quantity because $V_{\rm bul}$ declines with radius while $V_{\rm disk}$ and $V_{\rm gas}$ rise. For zC-400569 we find $M_{\rm bul}/M_{\rm bar}\simeq0.2$, comparable to late-type spirals (Sc or Sb), while for zC-488879 we find $M_{\rm bul}/M_{\rm bar}\simeq0.5$, comparable to early-type disks (Sa or S0).

We compare the stellar masses ($M_\star = M_{\rm bul} + M_{\rm disk}$) from our dynamical model with those independently measured from SED fitting using automated pipelines \citep{Liu2019}. For zC-400569, we find $M_\star = (7.8^{+2.3}_{-2.8})\times  10^{10}$ M$_\odot$ that is consistent with the SED value of $(21.9 \pm 6.5) \times 10^{10}$ M$_\odot$ at the 2$\sigma$ level. For zC-488879, we find $M_\star = (21.1^{+3.6}_{-4.2})\times10^{10}$ that is significantly different from the SED value of $(5.2\pm1.2) \times 10^{10}$ M$_\odot$ at the 3$\sigma$ level. The high stellar mass of this galaxy is dominated by the bulge component, which is unavoidably needed to explain the inner rotation speeds of $\sim$300 \kms. This indicates that stellar masses from automated SED fitting may occasionally be uncertain up to a factor of 4. 

In principle, stellar masses from SED fitting and/or galaxy colors may be improved by considering the bulge and stellar disk separately \citep[e.g.,][]{Schombert2022}. This requires spatially resolved images that are currently available only in the rest-frame UV and optical parts of the spectrum (see Sect.\,\ref{sec:HST}), which are most sensitive to the unavoidable assumptions on star-formation history, chemical enrichment, and dust extinction \citep[e.g.,][]{Schombert2019}. The situation may be improved with JWST images, probing the rest-frame NIR part of the spectrum, which is less sensitive to those assumptions and may therefore provide accurate stellar masses for the bulge and disk components separately.

As a consistency check, we compute an effective $\alpha_{\rm CO}=M_{\rm gas}/L_{\rm CO(1-0)}$ in units of M$_{\odot}$ (K\,\kms\,pc$^2$)$^{-1}$, using the gas mass from our dynamical model and the observed CO($2-1$) and CO($3-2$) luminosities with the assumption of average $R_{21}=0.85$ and $R_{32}=0.60$. We find $\alpha_{\rm CO} = 2.1^{+0.7}_{-0.8}$  for zC-400569 and $\alpha_{\rm CO} = 2.9^{+0.7}_{-0.9}$ M$_{\odot}$ for zC-488879. Both values are consistent with the Milky-Way $\alpha_{\rm CO}$ within the uncertainties and are within the range of values found in star-forming galaxies across cosmic time (see Table 14 in \citealt{Dunne2022}). For zC-400569, we make a similar consistency check for $\alpha_{\ci}=M_{\rm gas}/L_{\ci(1-0)}$ using the \ci($1-0$) emission, which is too faint for rotation-curve measurements but is a good global tracer of the H$_2$ mass \citep{Papadopoulos2004}. We find $\alpha_{\ci}=6.9^{+1.9}_{-2.4}$ which is in tension at more than 4$\sigma$ with the latest calibration of $17.0\pm0.3$ from \citet{Dunne2022} using a sample of 407 galaxies across cosmic time. Our value, however, is very close to $\alpha_{\ci}=7.3$ found by \citet{Crocker2019} in 18 nearby galaxies using spatially resolved, multitransition CO and \ci\ observations. It is also consistent with theoretical expectations from 3D astrochemical simulations \citep{Bisbas2021}.

\subsection{Mass models adding a dark matter halo}\label{sec:darkmatter}

Fig. \ref{fig:MassModels} (middle panels) shows mass models including a DM halo. From a statistical perspective, adding a DM halo with additional free parameters is not necessary because the rotation curves are already well fitted by the baryon-only mass model. The following models, however, allow us to check whether the existing data are consistent with expectations from $\Lambda$CDM cosmology.

We assume a spherical DM halo with a Navarro-Frenk-White (NFW) density profile \citep{Navarro1996}, which has two free parameters: the halo concentration $C_{200}$ and the halo mass $M_{200}$ (or equivalently the halo velocity $V_{200}$). These quantities are defined in the same way as in \citet{Li2020}. Since the observed rotation curve can be fully explained by baryons, the halo parameters cannot be constrained using only the kinematic data. Thus, we impose two $\Lambda$CDM scaling relations as Bayesian priors \citep[following][]{Li2020}: (1) the $M_\star-M_{200}$ relation from abundance$-$matching techniques, and (2) the $M_{200}-C_{200}$ relation from cosmological N-body simulations. Specifically, we impose the $M_\star-M_{200}$ relations derived by \citet{Legrand2019} at $z=[1.1-1.5]$ and $z=[2.0-2.5]$ on zC-488879 and zC-400569, respectively.
Similarly, the mass-concentration relations from \citet{Dutton2014} at $z\simeq1$ and $z\simeq2$ are imposed on zC-488879 and zC-400569, respectively. We also consider the redshift-dependent scatters given by \citet{Legrand2019} on $\log(M_{200})$ and by \citet{Dutton2014} on $\log(C_{200})$. Both scatters are on the order of 0.1 dex.

With respect to the baryons-only model, the MCMC fits decrease the gravitational contributions of stellar and gas disks to leave room for the DM contribution. The bulge contribution, instead, is almost unchanged given its characteristic declining shape, so the value of $M_{\rm bul}/M_{\rm bar}$ increases for both galaxies. For zC-400569, the best-fit masses decrease by a factor of $\sim$1.7 in stars and $\sim$2.1 in gas, while for zC-488879 they decrease by a factor of $\sim$1.4 in stars and $\sim$3.1 in gas. The actual uncertainties on stellar and gas masses of high-$z$ galaxies are surely as large as a factor of 3, so we cannot categorically rule out the scenario in which the DM halo is dynamically important in the inner galaxy regions ($R\lesssim 8-9$ kpc). Moreover, the best-fit parameters lie on the $M_\star-M_{200}$ and $M_{200}-C_{200}$ relations (imposed as priors) within the uncertainties. Thus, we conclude that the observed rotation curves are consistent with ``cuspy'' $\Lambda$CDM halos due to a severe disk-halo degeneracy, as we discuss in Sect.\,\ref{sec:reloaded}.

\subsection{Mass models in Milgromian dynamics}\label{sec:MOND}

Milgromian dynamics \citep[MOND;][]{Milgrom1983a, Milgrom1983b, Milgrom1983c} is the major alternative to particle DM. The MOND paradigm modifies the laws of gravity and/or inertia when accelerations are smaller than an acceleration scale $a_{\rm 0} \simeq 10^{-10}$ m\,s$^{-2}$, which is typical for the outer parts of spiral galaxies \citep[see][for reviews]{Famaey2012, Milgrom2014, Banik2022}.
The rotation curves of the two galaxies studied here are limited to the inner high-acceleration regions where $V_{\rm obs}^2/R > 3-4 a_0$, so they are expected to probe the Newtonian regime of the theory. Still, it is important to test MOND in these high-$z$ galaxies for various reasons: (1) the MOND acceleration scale may possibly vary with cosmic time; (2) the $\Lambda$CDM relation between angular distance ($D_{\rm A}$) and redshift may not apply in a MOND cosmology, so rotation curve fits of high-$z$ galaxies may provide empirical constraints on the $D_{\rm A}(z)$ relation in MOND; (3) the MOND interpolation function, linking the Newtonian and Milgromian regimes, may not be constant with cosmic time.

Constraining the redshift evolution of $a_0$ is particularly important \citep{Milgrom2017}. Empirically, it is known that $a_{0}\simeq c \times H_0$ and $a_{0}\simeq c^{2}\sqrt{\Lambda}$, where $c$ is the speed of light and $\Lambda$ is the cosmological constant. Both equivalences may be mere numerical coincidences, or they may have a deeper physical meaning. For example, the first coincidence may suggest that $a_0(z) = c \times H(z)$, while the second one that $a_{0}$ does not vary with $z$.

\begin{figure*}
\centering
\includegraphics[width=\textwidth]{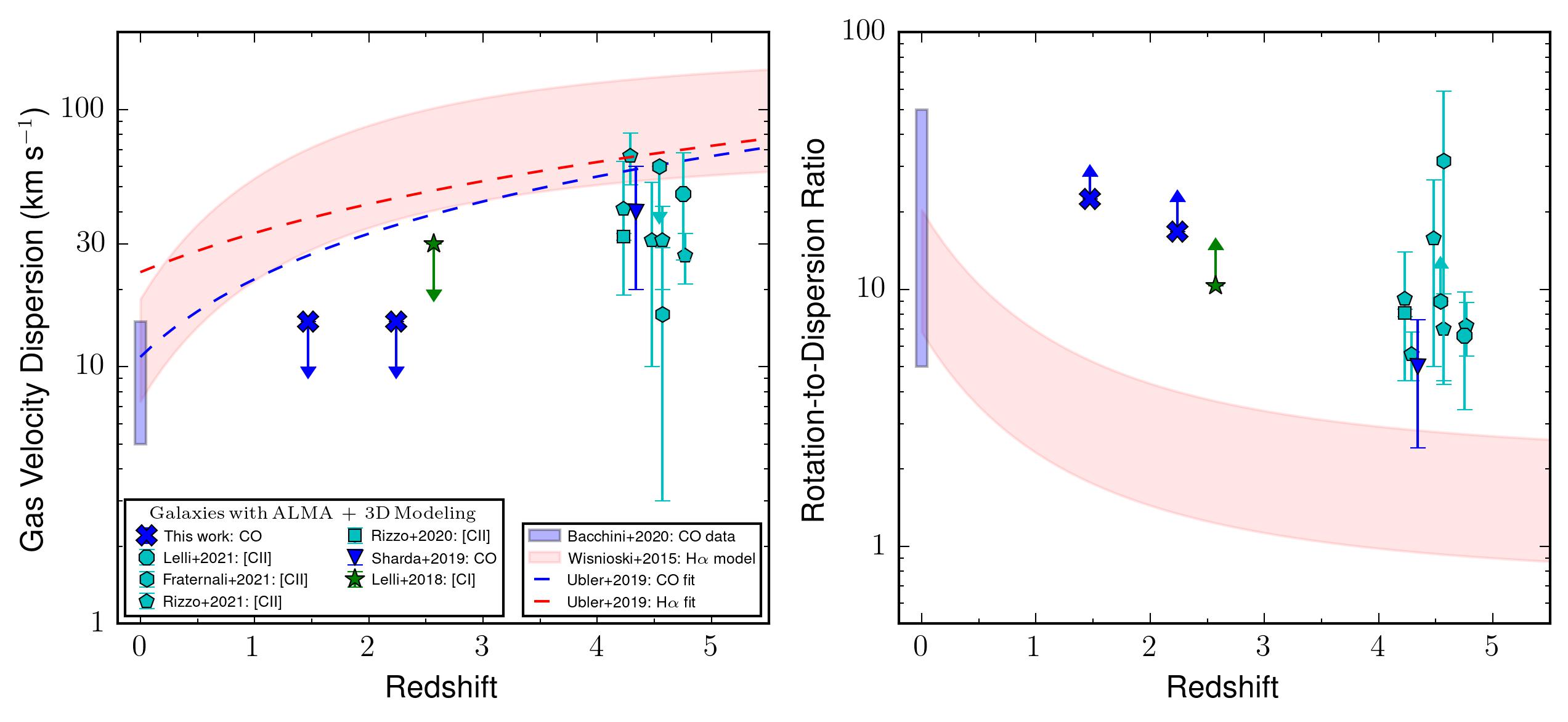}
\caption{Gas velocity dispersion (left panel) and $V_{\rm rot}/\sigma_{\rm V}$ ratio (right panel) as a function of redshift. The pink band shows the model from \citet{Wisnioski2015} based on the Toomre-disk instability criterion. The dashed lines show the best-fit functions from \citet{Ubler2019} using IFU data of ionized gas (red) and NOEMA data of molecular gas (blue). The points show individual galaxies that satisfy two quality criteria: (1) ALMA observations of cold gas tracers such as CO (blue points), \ci\ (green), and \cii\ (cyan) lines, and (2) detailed kinematic modeling with \bb.}
\label{fig:Turbulence}
\end{figure*}

We perform MOND fits assuming (1) the empirical value $a_0 = 1.2 \times 10^{-10}$ m s$^{-2}$ found at $z=0$ \citep[e.g.,][]{Lelli2022}, (2) the theoretical $D_{\rm A}-z$ relation from $\Lambda$CDM, which is expected to hold in some relativistic MOND theories \citep{Skordis2021}, and (3) the interpolation function that best fits the empirical radial acceleration relation (RAR) at $z\simeq0$ \citep{McGaugh2016, Lelli2017}. Figure\,\ref{fig:MassModels} (right panels) shows the mass models from MOND fits. With respect to the baryon-only Newtonian mass models, the stellar and gas masses are decreased by $10\%-30\%$ due to a moderate MOND effect at large radii. The resulting values of $M_\star$ and $M_{\rm gas}$ are physically acceptable as discussed in Sect.\,\ref{sec:onlybaryons}. This implies that rotation curves at $z\simeq1-2$ are compatible with the value of $a_0$ and the interpolation function measured at $z\simeq0$.

\section{Discussion}

\subsection{Rotation versus pressure support in high-$z$ galaxies}

In Sect.\,\ref{sec:disp} we find that the CO velocity dispersion of both zC-400569 and zC-488879 is surprisingly low. We cannot robustly measure $\sigma_{\rm CO}$ because the observed line broadening is fully dominated by the effects of spatial and spectral resolution, so we have a fiducial upper limit of 15 \kms set by the spectral resolution. The resulting lower limits in the $V_{\rm rot}/\sigma_{\rm CO}$ ratio are on the order of 17-22, so the molecular gas disk is fully supported by rotation and there is no need of corrections for pressure support. These properties are comparable to those of spiral galaxies at $z\simeq0$: local CO disks typically have $\sigma_{\rm CO}\simeq5-15$ \kms \citep{Mogotsi2016, Bacchini2020} and $V_{\rm rot}/\sigma_{\rm CO} \simeq 5-50$, with lower $V_{\rm rot}/\sigma_{\rm CO}$ values at small radii and higher $V_{\rm rot}/\sigma_{\rm CO}$ values in the outer parts \citep[see Appendix in][]{Bacchini2020}. In addition, ALMA observations at exceptionally high spatial resolutions ($10-30$ pc) revealed that local galaxies can host nuclear CO disks with extremely low velocity dispersions of just $1-2$ \kms\ and $V_{\rm rot}/\sigma_{\rm CO} \gtrsim 100$ \citep{Davis2017, Davis2018}. This fact hints at the possibility that, even at $z\simeq0$, velocity dispersions from CO data with moderate spatial ($\sim$500 pc) and spectral ($\sim$5 \kms) resolutions may be inflated by observational effects. Evidently, measuring the intrinsic velocity dispersion of a rotating disk is not a trivial exercise.

Figure\,\ref{fig:Turbulence} (left panel) shows the redshift evolution of $\sigma_{\rm V}$ inferred by \citet{Ubler2019} using both IFU data of ionized gas and NOEMA data of molecular gas. We also show the prediction of a toy model based on the Toomre-instability criterion, which was built by \citet{Wisnioski2015} to reproduce ionized gas data at $z\lesssim3.5$. For comparison, we consider a small sample of galaxies that satisfy two quality criteria: (1) high-resolution ALMA data of cold gas tracers such as CO, \ci, and \cii\ lines, and (2) detailed 3D kinematic modeling for consistency with the current work. In addition to the two main-sequence galaxies studied here, this sample include an AGN-host galaxy at $z\simeq2.6$ with \ci($2-1$) data \citep{Lelli2018} and several submillimeter galaxies at $z\simeq4-5$ with \cii\ data, considering both lensed \citep{Rizzo2020, Rizzo2021} and unlensed \citep{Sharda2019, Lelli2021, Fraternali2021} sources. The velocity dispersions from ALMA data appear to be systematically below the expectations from IFU data of ionized gas as well as NOEMA data of molecular gas. Figure\,\ref{fig:Turbulence} (right panel) compares the $V_{\rm rot}/\sigma_{\rm V}$ ratios of this sample with the model of \citet{Wisnioski2015}. The $V_{\rm rot}/\sigma_{\rm V}$ of galaxies with ALMA data are higher than expected from the extrapolation of the model, but comparable to those of rotation-supported spiral galaxies at $z\simeq0$. 

In our view, it is possible that previous measurements from IFU and/or NOEMA data may have not been fully corrected for beam-smearing effects, leading to systematic overestimates of $\sigma_{\rm V}$ and underestimates of $V_{\rm rot}/\sigma_{\rm V}$ \citep[see also][]{DiTeodoro2016}. In fact, as the data quality increases, the measured velocity dispersion decreases \citep[see Fig. 6 in][]{Rizzo2021}. In addition, ionized gas tracers may be more easily ``contaminated'' by large-scale wind components than $\rm H_2$-only tracers, leading to an artificial increase of the measured velocity dispersion \citep[see the discussion in][]{Lelli2018}. The \hi-to-H$_2$ phase-transition criterion, indeed, is much more readily satisfied in the high-pressure regions inside the star-forming disks than in the wind regions outside them (though in some AGN and/or starburst galaxies, CO-rich $\rm H_2$ winds do exist). Moreover, a CO disk would correspond to lower disk scale-heights $z_{\rm disk}$ and thus smaller $\sigma_{\rm V}$ because in the vertical direction one has $\sigma_{\rm V} \simeq z_{\rm disk} \sqrt{2\pi G \langle\rho_0\rangle}$, where $\langle\rho_0\rangle$ is the average mid-plane mass density. In conclusion, it could well be that high-$z$ disks are not as turbulent as generally thought, but more high-resolution and high-sensitivity ALMA observations are needed to confirm or refuse this possibility.

\subsection{Rotation-curve shapes: Inner galaxy regions}

In Sect.\,\ref{sec:rotcur}, we find that the rotation curves of zC-400569 and zC-488879 are flat with no sign of decline out to $R\simeq8$ kpc. Different emission lines, such as CO($2-1$), CO($3-2$), CO($4-3$) and H$\alpha$, provide consistent results (Fig.\,\ref{fig:rotcur}). This fact suggests that the various gas phases are kinematically settled and that the resulting rotation curve probes the equilibrium gravitational potential.

Flat rotation curves have been previously found in galaxies at $z\simeq1$ using H$\alpha$ kinematics from IFU data \citep{DiTeodoro2016, Sharma2021}. Specifically, \citet{DiTeodoro2016} analyzed a carefully selected sample of 18 galaxies with high-quality H$\alpha$ data, while \citet{Sharma2021} studied a statistical sample of 344 galaxies with data of variable quality. Both studies used nonparametric tilted-ring modeling, so they provide the actual rotation curve on a ring by ring basis, not a functional parametrization thereof. Using parametric models, instead, \citet{Genzel2017} studied six galaxies at $z\simeq0.8$ to 2.4 (including zC-400569) and found declining rotation curves at large radii. Subsequently, \citet{Genzel2020} increased their sample to 41 galaxies (including CO data from NOEMA for seven of them) and found a variety of rotation-curve shapes, from rising, to flat, to declining. A similar variety in rotation-curve shapes is found by \citet{Bouche2022} for nine galaxies at $z\simeq1$, fitting parametric models to \oii\ observations from MUSE.

These different results may be driven by the different modeling techniques, but they could also indicate a genuine variety among galaxies. At $z\simeq0$ it is well established that the \textit{inner} rotation curves ($R\lesssim 2-3 R_{\rm e}$) display a wide variety of shapes, depending on the galaxy morphology, surface brightness, and bulge-to-disk ratio \citep[e.g.,][]{Corradi1990, Casertano1991, deBlok1996, Tully1997, Noordermeer2007, Swaters2009, Lelli2022}. This evidence is nicely summarized by the so-called Renzo's rule: ``For any feature in the luminosity profile of a galaxy there is a corresponding feature in the rotation curve, and vice versa'' \citep{Sancisi2004}. Quantitatively, the inner steepness of the rotation curve, proxy of the central dynamical surface density, closely correlates with the central surface brightness, proxy of the central baryonic surface density \citep{Lelli2013, Lelli2014b, Lelli2016c}. Thus, it may not be surprising that a similar diversity in the \textit{inner} rotation curves exists at cosmic noon, given the observed diversity in the light (or stellar mass) distribution of galaxies.

\begin{figure}
\centering
\includegraphics[width=0.49\textwidth]{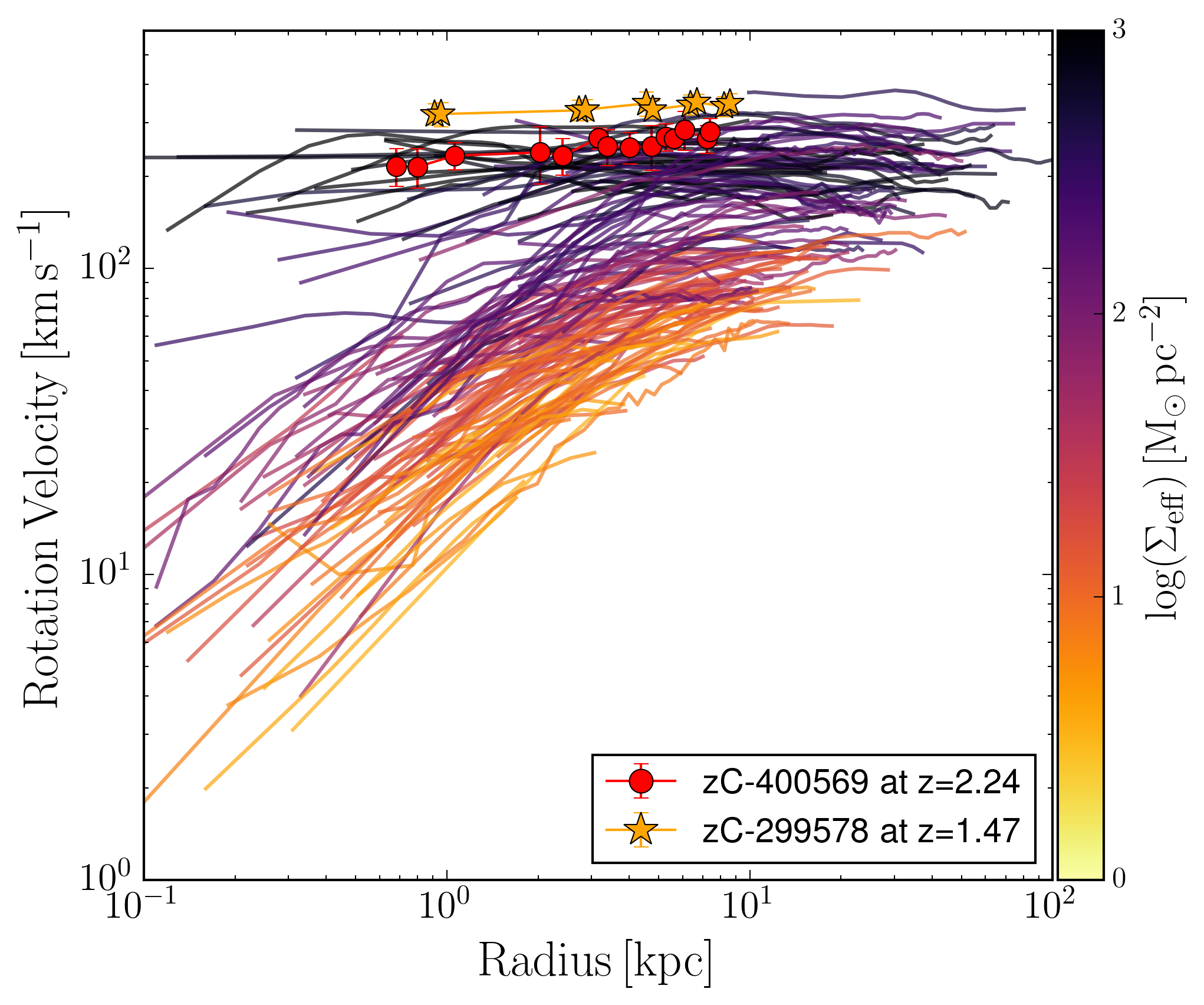}
\caption{The rotation curves of of zC-400569 at $z\simeq2.24$ (red dots) and zC-488879 at $z\simeq1.47$ (orange stars) are compared with those of disk galaxies at $z\simeq0$ from the SPARC database \citep{Lelli2016b}. The latter ones are color-coded by the mean stellar surface density within the effective radius. Rotation curves at $z\simeq0$ are much more extended than those at high $z$ thanks to the availability of \hi\ observations.}
\label{fig:SPARC}
\end{figure}

\subsection{Rotation-curve shapes: Outer galaxy regions}

The vast majority (if not all) of rotation curves at $z\simeq0$ become flat at very large radii, outside the bright stellar component of galaxies, which can be probed with \hi\ observations \citep{Bosma1981, vanAlbada1985, Begeman1991, Sanders1998b}. To illustrate the situation, Figure\,\ref{fig:SPARC} compares the rotation curves of zC-400569 and zC-488879 with those of local galaxies from the SPARC sample \citep{Lelli2016b}. The \hi\ rotation curves of massive spirals at $z\simeq0$ reach radii out to $50-100$ kpc, so they are much more extended than the CO or H$\alpha$ rotation curves of galaxies at cosmic noon, halting at $R\simeq8$ kpc. Unfortunately, the \hi\ line is weak and cannot be detected (nor spatially resolved) in individual galaxies at $z\simeq1-3$, probably not even with the future Square Kilometre Array at full capacity.

We stress that zC-400569 is one of the galaxy with the best IFU data \citep{Genzel2017}, so represents the current state of the art in terms of H$\alpha$ kinematics at cosmic noon. In principle, to increase the sensitivity in the outer galaxy regions, one may stack the line emission of multiple galaxies after normalizing both the radial extent and the velocity extent in some way. Such stacking experiments have been performed \citep{Lang2017, Tiley2019} but led to controversial results because the final stacked rotation curves depend on the adopted normalization, which is not surprising given the observed variety in the inner rotation-curve shapes of individual galaxies.

Generally speaking, H$\alpha$ and CO emissions are cospatial with the star-forming disk, so they cannot easily probe the ubiquitous flattening of rotation curves beyond the bright stellar component of galaxies. This problem is even more prominent for the high-J CO lines (J=$3-2$ and higher) that are routinely imaged by ALMA at high $z$. In fact, high-J CO lines trace only dense and warm H$_2$ gas that is closely associated with active star-forming sites, unlike the lower-J CO lines that trace also less dense and cooler H$_2$ phases. A possible alternative is the $S(0):\,{\rm J_u-J_l}=2-0$ rotational line of H$_2$ at 28\,$\mu$m that emanates from a warm molecular gas phase concomitant with the cold atomic phase in spirals \citep{Papadopoulos2002}, but there is currently no telescope to image this line in galaxies across cosmic time. The most promising line at high $z$ is the bright \cii\ emission at 158$\mu$m, which traces a combination of atomic, molecular, and ionized gas, but this line is most easily observed by ALMA only at $z>4$ \citep[e.g.,][]{Lelli2021}.

Despite all these limitations, Figure\,\ref{fig:SPARC} shows a remarkable fact: the inner rotation curves ($R<8-9$ kpc) of zC-400569 and zC-488879 are very similar to those of local galaxies with similar stellar surface densities ($\sim$10$^{3}$ M$_\odot$ pc$^{-2}$). This result suggests that main-sequence galaxies at cosmic noon are dynamically evolved and may turn into local spiral galaxies without the need for a major redistribution of their total mass, at least within the inner 10 kpc. This fact is in line with ALMA kinematic studies of galaxies at $z>4$ \citep{Rizzo2020, Rizzo2021, Lelli2021, Tsuki2021} and recent results at $z>3$ \citep{Ferreira2022} from the James Webb Space Telescope (JWST), pointing to speedy galaxy evolution in the first billions year of the Universe's lifetime.

\subsection{The dark matter effect: Probing low accelerations}

In Sect.\,\ref{sec:onlybaryons}, we find that the rotation curves of both zC-400569 and zC-488879 can be fitted with a mass model that contains only baryonic components. This occurs even though the rotation curve is flat due to the sum of a declining bulge contribution at small radii and a rising disk contribution in the outer parts (see Fig.\,\ref{fig:MassModels}). Similar mass models are common for local spirals \citep{Lelli2016b}. Historically, indeed, the evidence for the DM effect in galaxies did not come simply from ``flat rotation curves'' because most H$\alpha$ rotation curves can be fitted by mass models with only baryonic components \citep{Kalnajs1983, Kent1986}. The unambiguous evidence for the DM effect came from rotation curves that \textit{remain} flat outside the stellar component, beyond the peak of the disk contribution at $2-3 R_{\rm d}$, such as those from deep \hi\ observations (\citealt{vanAlbada1985, Kent1987, Begeman1989}; see \citet{Sanders2014} for an  historical perspective).

\begin{figure}
\centering
\includegraphics[width=0.48\textwidth]{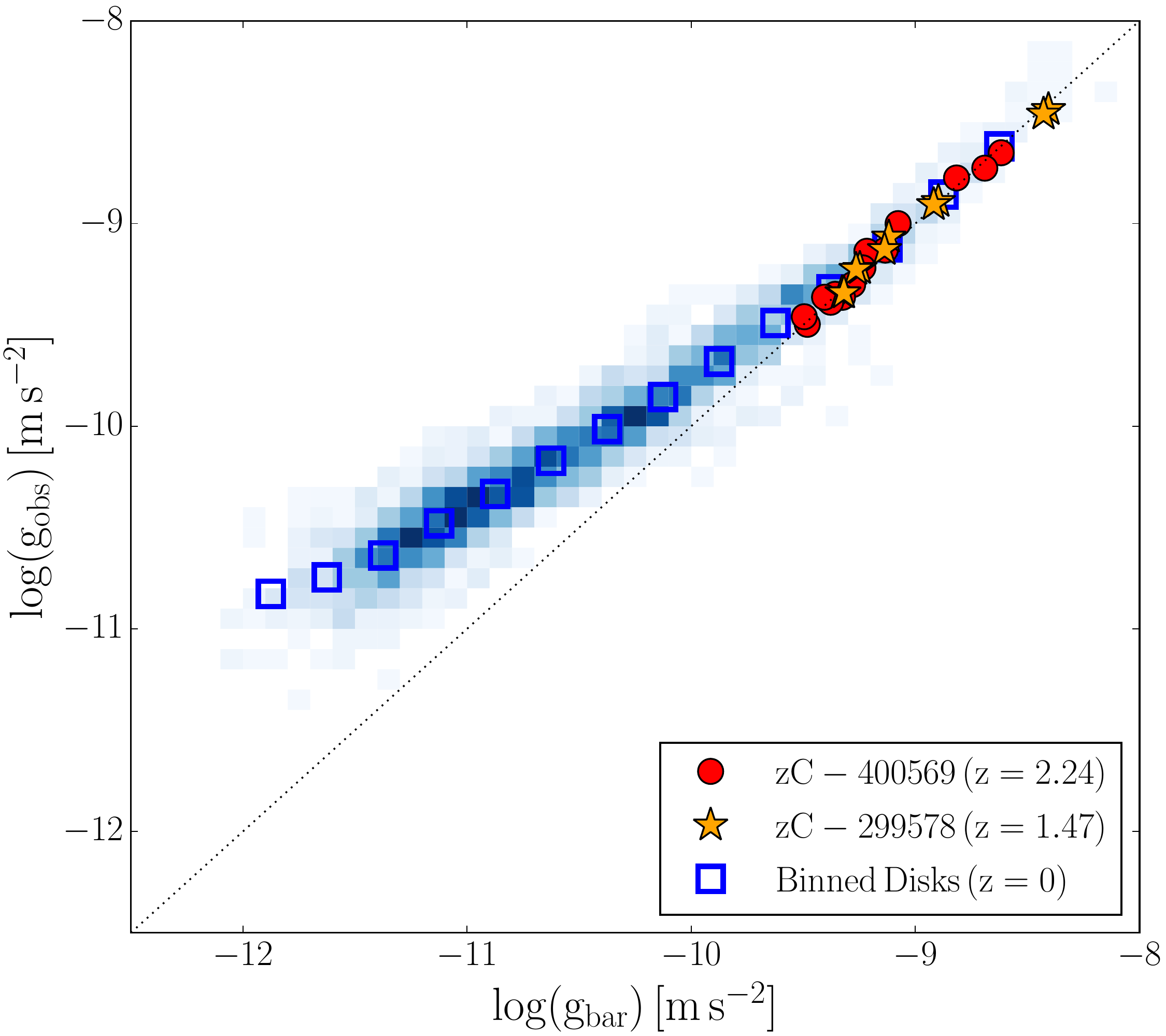}
\caption{Location of zC-400569 (red dots) and zC-488879 (orange stars) on the radial acceleration relation at $z\simeq0$ (blue colorscale and open squares). For both galaxies we assume the values of $\Upsilon_{\rm gas}$, $\Upsilon_{\rm disk}$, and $\Upsilon_{\rm bul}$ from the baryons-only mass model (Sect.\,\ref{sec:onlybaryons}). Both galaxies lie on the 1:1 relation (dotted line) at high accelerations, similarly to disk galaxies at $z\simeq0$. To unambiguously probe the DM effect ($g_{\rm obs}>g_{\rm bar}$) one needs more extended rotation curves probing lower accelerations.}
\label{fig:RAR}
\end{figure}

To illustrate the situation, Figure\,\ref{fig:RAR} shows the location of zC-400569 and zC-299569 on the empirical RAR at $z\simeq0$ \citep{McGaugh2016, Lelli2017}: the observed centripetal acceleration from rotation curves ($g_{\rm obs}=V_{\rm rot}^2/R$) is plotted against that expected from the distribution of baryons ($g_{\rm bar} = -\nabla \phi_{\rm bar}$). For local galaxies, the value of $g_{\rm bar}$ is well measured using Spitzer photometry at 3.6 $\mu$m, which provides a dust-free tracer of the stellar mass distribution and allows for the adoption of a common mass-to-light ratio for all galaxies \citep[e.g.,][]{Lelli2017}. For high-$z$ galaxies, rest-frame NIR photometry is not available yet, so we need to make assumptions on stellar masses to locate them on the RAR. This is similar to the observational situation at $z\simeq0$ about 20 years ago \citep{McGaugh2004}. 

In Figure\,\ref{fig:RAR}, we assume the values of $\Upsilon_{\rm disk}$, $\Upsilon_{\rm bul}$, $\Upsilon_{\rm disk}$ from the baryon-only mass model (Sect.\,\ref{sec:onlybaryons}). As expected, both high-$z$ galaxies lie on the unity line ($g_{\rm obs} = g_{\rm bar}$) at accelerations higher than a few 10$^{-9}$ m\,s$^{-2}$. Clearly, lower values of $\Upsilon_{\rm disk}$, $\Upsilon_{\rm bul}$, $\Upsilon_{\rm disk}$ (such as those found in the NFW model, Sect.\,\ref{sec:darkmatter}) would shift $g_{\rm bar}$ toward the left and move the data above the unity line. In local galaxies, the DM effect ($g_{\rm obs}>g_{\rm bar}$) appears at accelerations lower than $\sim$10$^{-10}$ m\,s$^{-2}$. If we assume that the rotation curves of our high-$z$ galaxies remain flat at the observed values, then they should be traced out to $20-30$ kpc to unambiguougly probe the DM-dominated regime. Unfortunately, these distances seem out of reach for H$\alpha$ and CO data from existing facilities.

\subsection{The disk-halo degeneracy reloaded}\label{sec:reloaded}

In Sect.\,\ref{sec:darkmatter}, we find that the rotation curves of both zC-400569 and zC-488879 can be fitted with a NFW halo in a $\Lambda$CDM context. Adding an NFW halo comes at the expense of decreasing the disk mass by a factor of $2-3$ with respect to the baryon-only model. As a result, for the NFW model, the DM contribution is already important in the inner galaxy regions. Given the large uncertainties on stellar and gas masses at $z\simeq1-3$, a NFW halo cannot be categorically ruled out.

The current situation is reminiscent of the well-known ``disk-halo degeneracy'' \citep{vanAlbada1985}, which has plagued the mass modeling of galaxies for a few decades. Essentially, the halo and disk contributions are degenerate because they have a similar shape (see Fig.\,\ref{fig:MassModels}), so one needs to know the disk mass with high accuracy to determine the DM halo properties and inner DM fractions. One way to break this degeneracy is using rest-frame NIR photometry \citep[e.g.,][]{Sanders1998b, Lelli2016b} because the stellar mass-to-light ratio is almost insensitive to the galaxy star-formation history, metallicity, and dust extinction \citep[e.g.,][]{Schombert2019}, so the stellar masses can be estimated with an accuracy of $\sim$25$\%$. This approach gives negligible DM fractions in the inner regions of massive galaxies at $z\simeq0$ (see Fig.\,\ref{fig:RAR}). For galaxies at cosmic noon, rest-frame NIR imaging will become available soon thanks to JWST, so we will be able to break the disk-halo degeneracy at $z\simeq1-3$. 

Our conclusions differ from those of \citet{Genzel2020} and \citet{Price2021}, who report high baryonic fractions and cored DM distributions in galaxies at cosmic noon \citep[see also][]{Bouche2022}.
Possibly, their findings are driven by an underrating of the disk-halo degeneracy, which may occur when the DM halo concentration is fixed and/or strong priors on stellar and gas masses are imposed. The most conservative conclusion is that it is just impossible to draw firm statements on the amount and distribution of DM in galaxies at high $z$ with the available data. However, the various similarities between our galaxies at $z\simeq1-2$ and massive spirals at $z\simeq0$ suggest that inner baryon dominance and core DM halos may, indeed, be the most likely scenario for galaxies at the cosmic noon. This scenario may soon be tested using rest-frame NIR images from JWST.

\section{Conclusions}

We presented high-resolution ALMA observations of multiple CO transitions for two main-sequence galaxies at cosmic noon: zC-400569 at $z\simeq2.24$ and zC-488879 at $z\simeq1.47$. We also reanalyzed H$\alpha$ data for zC-400569 from SINFONI at the VLT. Our main results can be summarized as follows:

\begin{enumerate}
    \item Both galaxies have regularly rotating CO disks at $R\lesssim8$ kpc and show possible hints of minor mergers and/or mild interactions at larger radii;
    \item The rotation curves of both galaxies are flat out to $R\simeq8$ kpc and show no sign of a decline in the outer parts;
    \item The CO velocity dispersion seems low ($\sigma_{\rm CO} \lesssim 15$ \kms), smaller than the spectral resolution, so the gas disks have $V_{\rm rot}/\sigma_{\rm CO}\gtrsim 17-22$ indicating a low turbulence environment;
    \item Mass models from HST images reveal a severe disk-halo degeneracy: models with inner baryon dominance and models with a NFW halo can fit the observed rotation curves equally well with acceptable stellar and gas masses;
    \item Mass models in Milgromian dynamics (MOND) can fit the observed rotation curves with the same acceleration scale and interpolation function measured at $z\simeq0$.
\end{enumerate}

In the near future, rest-frame NIR imaging from JWST will provide accurate stellar masses (comparable to those at $z\simeq0$) and allow us to distinguish between various mass models, breaking the disk-halo degeneracy. In particular, we may be able to discriminate between $\Lambda$CDM models, which predict a significant DM effect within 8 kpc, and MOND models, which do not.

In terms of galaxy evolution, the dynamical properties of these two galaxies (rotation-curve shapes, high $V_{\rm rot}/\sigma_{\rm V}$ ratios, and bulge-to-total ratios) are remarkably similar to those of spiral galaxies at $z\simeq0$. This fact suggests that some massive galaxies at cosmic noon undergo weak dynamical changes over more than 10 Gyr of the Universe's lifetime, evolving into local spiral galaxies without the need of a severe mass redistribution in their inner regions.

\begin{acknowledgements}
Z.-Y.Z. and L.L. acknowledge the support of the National Natural Science Foundation of China (NSFC) under grants No. 12041305, 12173016. Z.-Y.Z. and L.L. acknowledge the science research grants from the China Manned Space Project with NOs.CMS-CSST-2021-A08 and CMS-CSST-2021-A07. Z.-Y.Z. and L.L. acknowledge the Program for Innovative Talents, Entrepreneur in Jiangsu. T.G.B. acknowledges support from Deutsche Forschungsgemeinschaft (DFG) grant no. 424563772.
This paper makes use of the following ALMA data: ADS/JAO.ALMA$\#$2017.1.01020.S and ADS/JAO.ALMA$\#$2019.1.00862. ALMA is a partnership of ESO (representing its member states), NSF (USA) and NINS (Japan), together with NRC (Canada), MOST and ASIAA (Taiwan), and KASI (Republic of Korea), in cooperation with the Republic of Chile. The Joint ALMA Observatory is operated by ESO, AUI/NRAO and NAOJ. Based on observations made with the NASA/ESA Hubble Space Telescope, and obtained from the Hubble Legacy Archive, which is a collaboration between the Space Telescope Science Institute (STScI/NASA), the Space Telescope European Coordinating Facility (ST-ECF/ESA) and the Canadian Astronomy Data Centre (CADC/NRC/CSA). 
\end{acknowledgements}

\bibliographystyle{aa} % style aa.bst
\bibliography{highz} % your references Yourfile.bib

\begin{appendix}

\section{Parameter space of kinematic models}\label{app:3Dfits}

\begin{figure*}
\centering
\includegraphics[width=\textwidth]{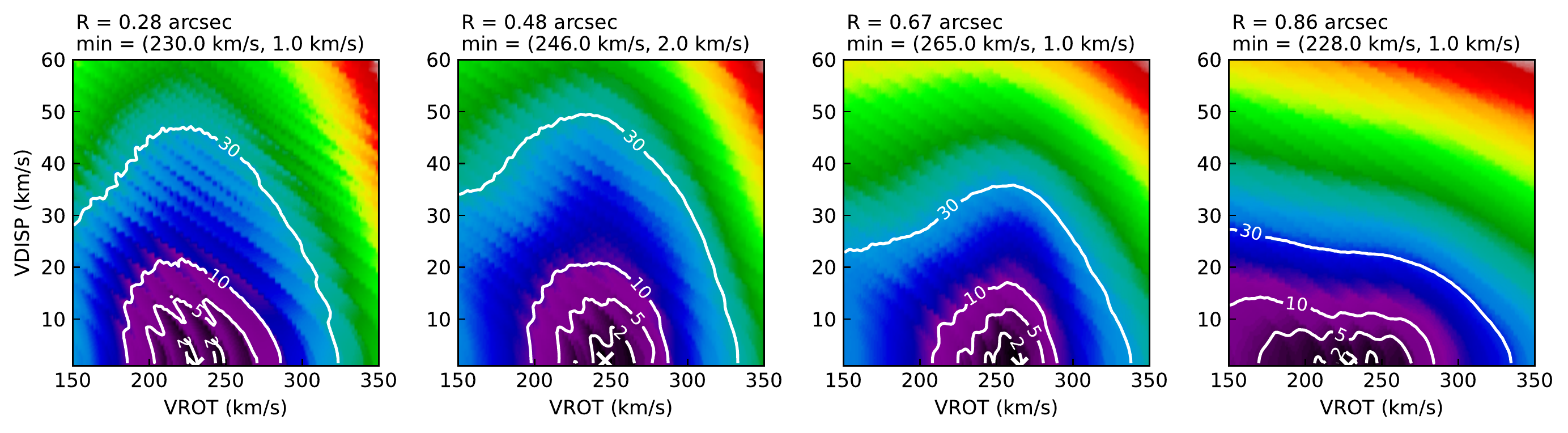}
\includegraphics[width=0.76\textwidth]{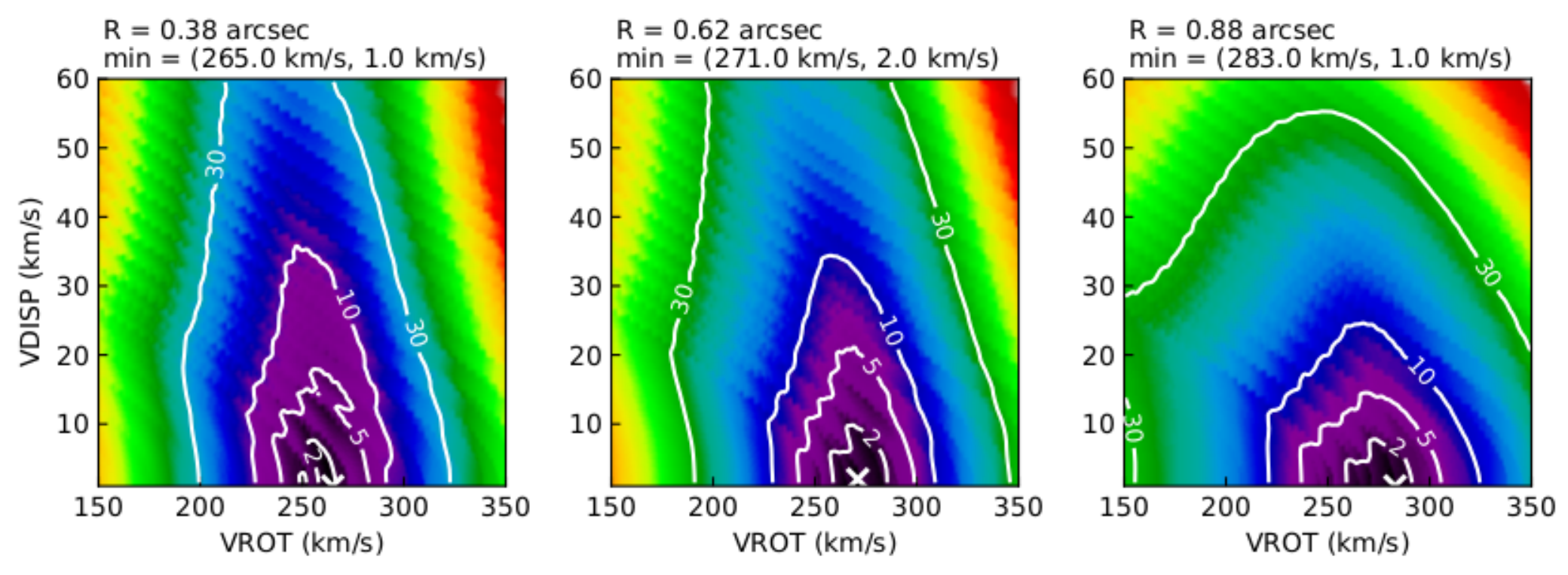}
\caption{Same as Figure \ref{fig:spacepar} but for the CO($3-2$) cube (top panels) and CO($4-3$) cube (bottom panels) of zC-400569.}
\label{fig:spacepar_bird1}
\end{figure*}

\begin{figure*}
\centering
\includegraphics[width=\textwidth]{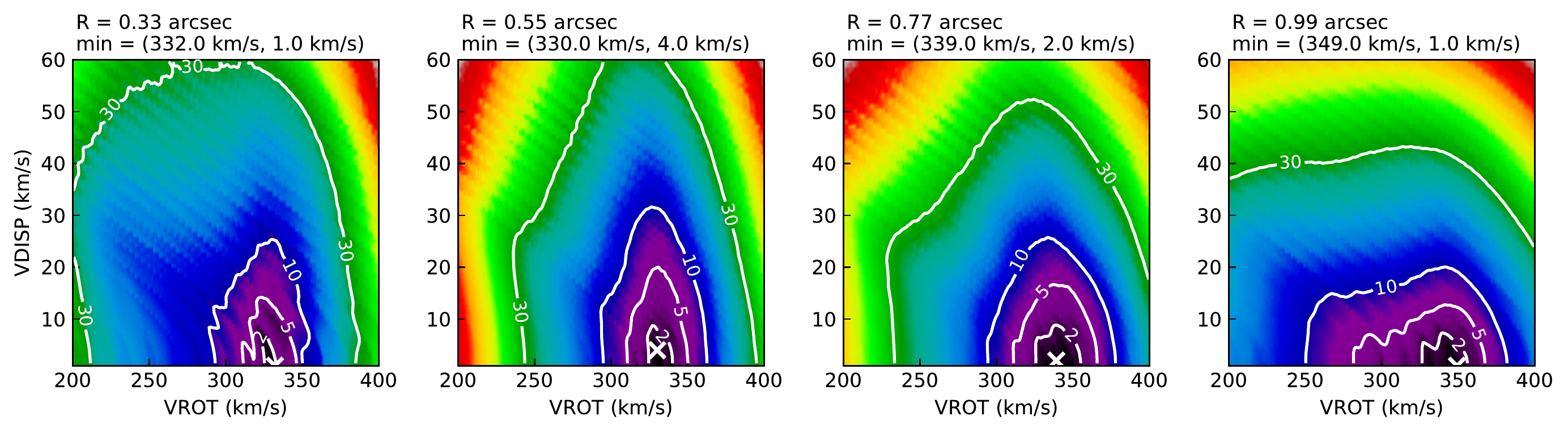}
\includegraphics[width=\textwidth]{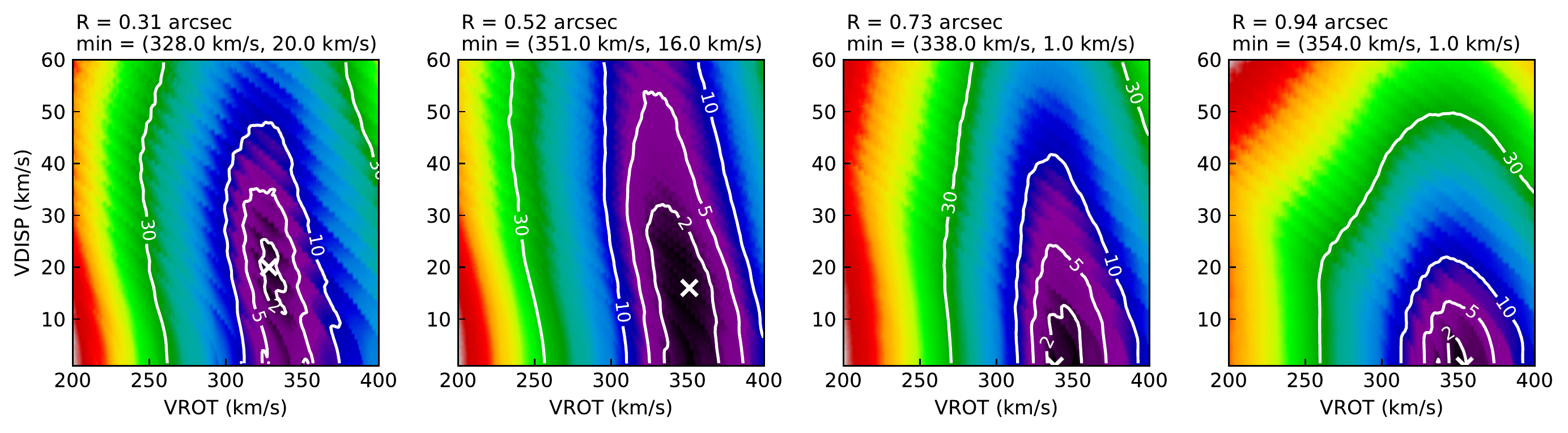}
\caption{Same as Figure\,\ref{fig:spacepar} but for the CO($3-2$) cube (top panels) and CO($4-3$) cube (bottom panels) of zC-299568.}
\label{fig:spacepar_bird2}
\end{figure*}

In Sect.\,\ref{sec:disp}, we used the task \texttt{SpacePar} of \bb\ to explore the $V_{\rm rot}-\sigma_{\rm V}$ parameter space, and showed the results for the H$\alpha$ cube of zC-400569 (Fig.\,\ref{fig:spacepar}). Here we present the results for the CO cubes of zC-500569 (Fig.\,\ref{fig:spacepar_bird1}) and zC-488879 (Fig.\,\ref{fig:spacepar_bird2}). For all emission lines, we exclude the innermost ring in which the beam-smearing effects are most severe. Fig.\,\ref{fig:spacepar_bird1} and Fig.\,\ref{fig:spacepar_bird2} show that the global minimum in $\sigma_{\rm V}$ occurs at very low values of $1-2$ \kms, suggesting that the intrinsic gas velocity dispersion cannot be reliably measured (see Sect.\,\ref{sec:disp} for details).

\section{Posterior probability distributions of mass models}\label{app:corner}

\begin{figure*}
\centering
\includegraphics[width=0.49\textwidth]{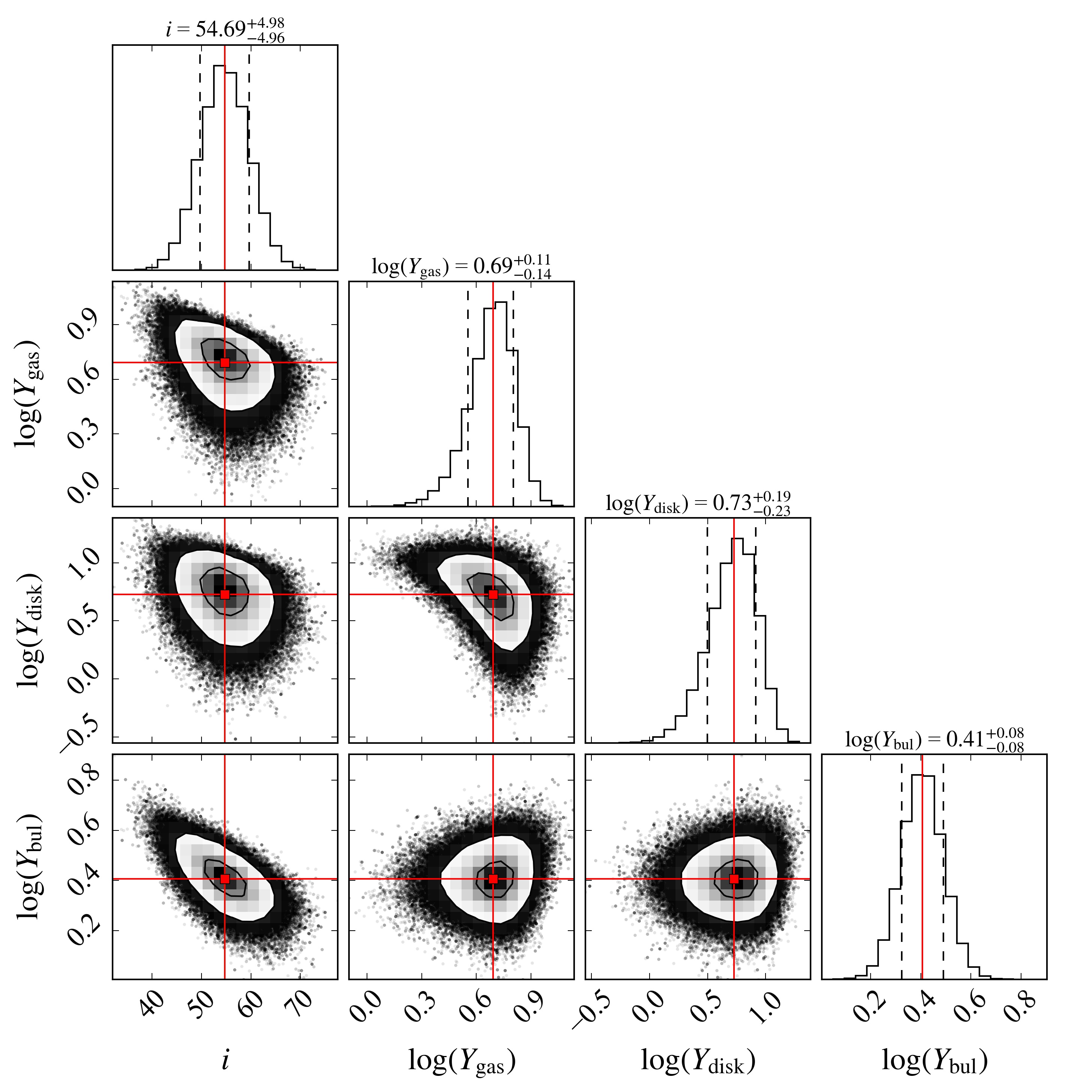}
\includegraphics[width=0.49\textwidth]{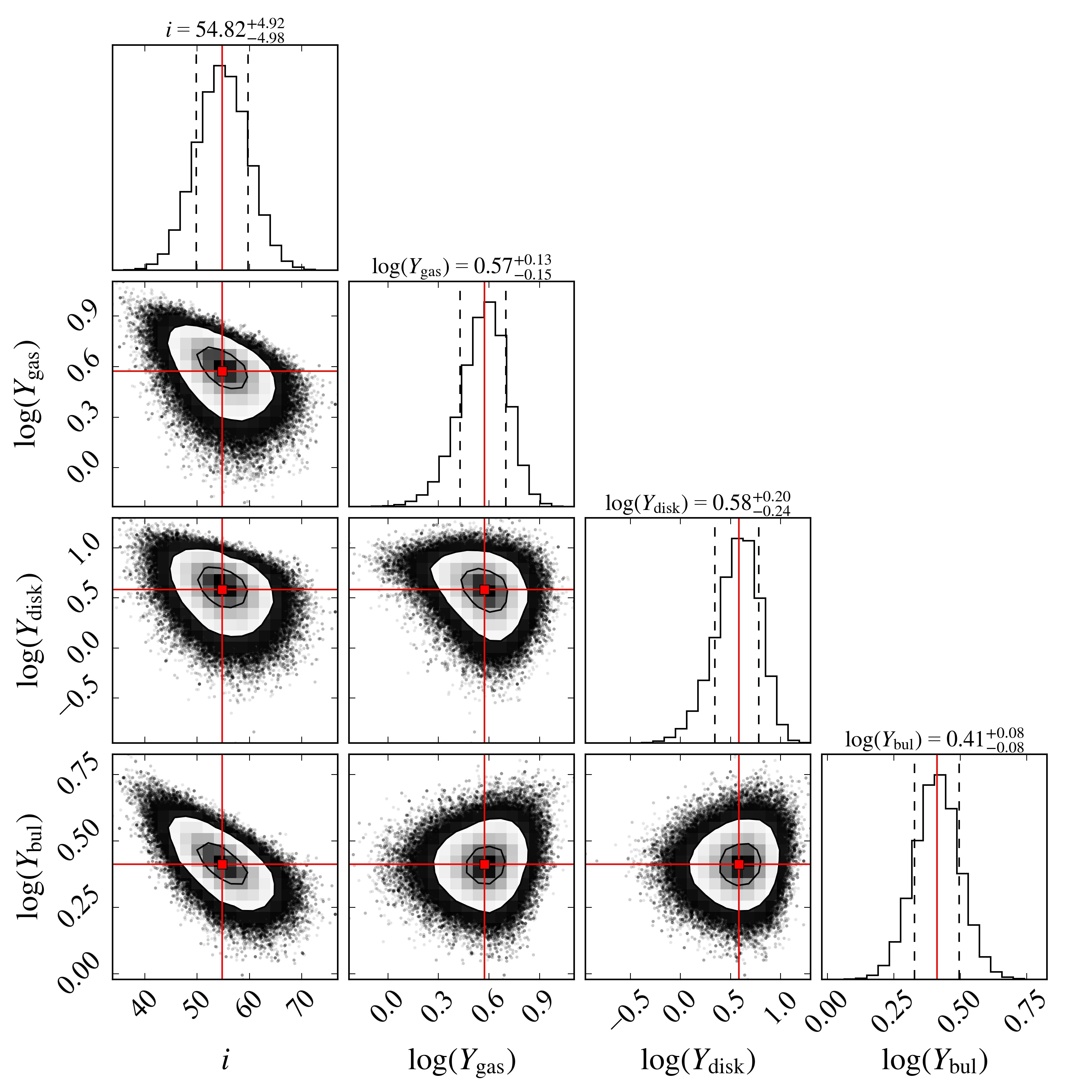}
\includegraphics[width=0.72\textwidth]{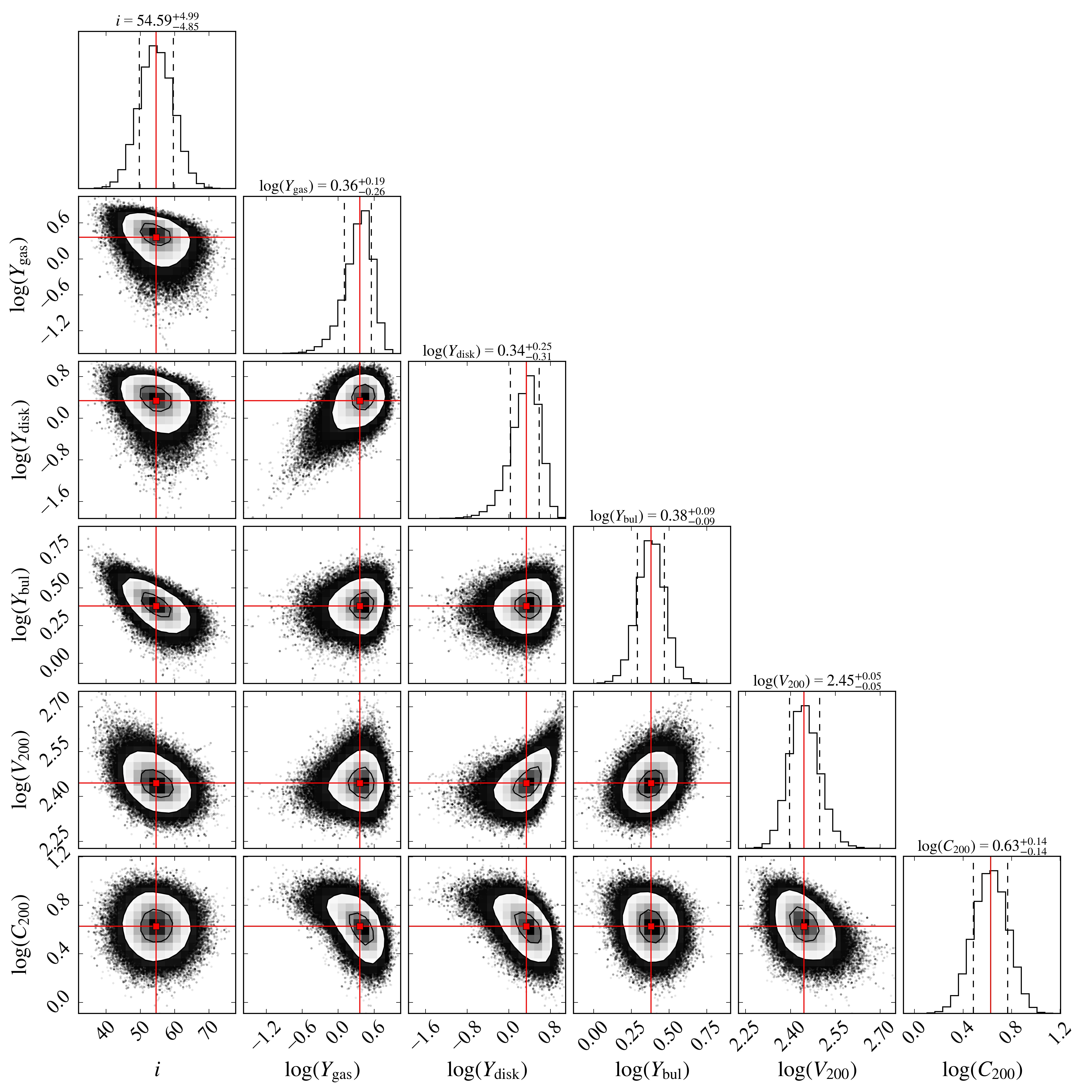}
\caption{Corner plots for different mass models of zC-400569 (see Sect.\,\ref{sec:massmodels}): baryons only (top left), MOND (top right), and baryons plus a NFW halo (bottom). The various panels show the posterior probability distribution of pairs of fitting parameters, and the marginalized probability distribution of each fitting parameter. In the inner panels, individual MCMC samples outside the 2$\sigma$ confidence region are shown with black dots, while binned MCMC samples inside the 2$\sigma$ confidence region are shown by a grayscale; black contours correspond to the 1$\sigma$ and 2$\sigma$ confidence regions; the red squares and solid lines show median values. In the outer panels (histograms), solid and dashed lines correspond to the median and $\pm1\sigma$ values, respectively.}
\label{fig:CornerBird1}
\end{figure*}

\begin{figure*}
\centering
\includegraphics[width=0.49\textwidth]{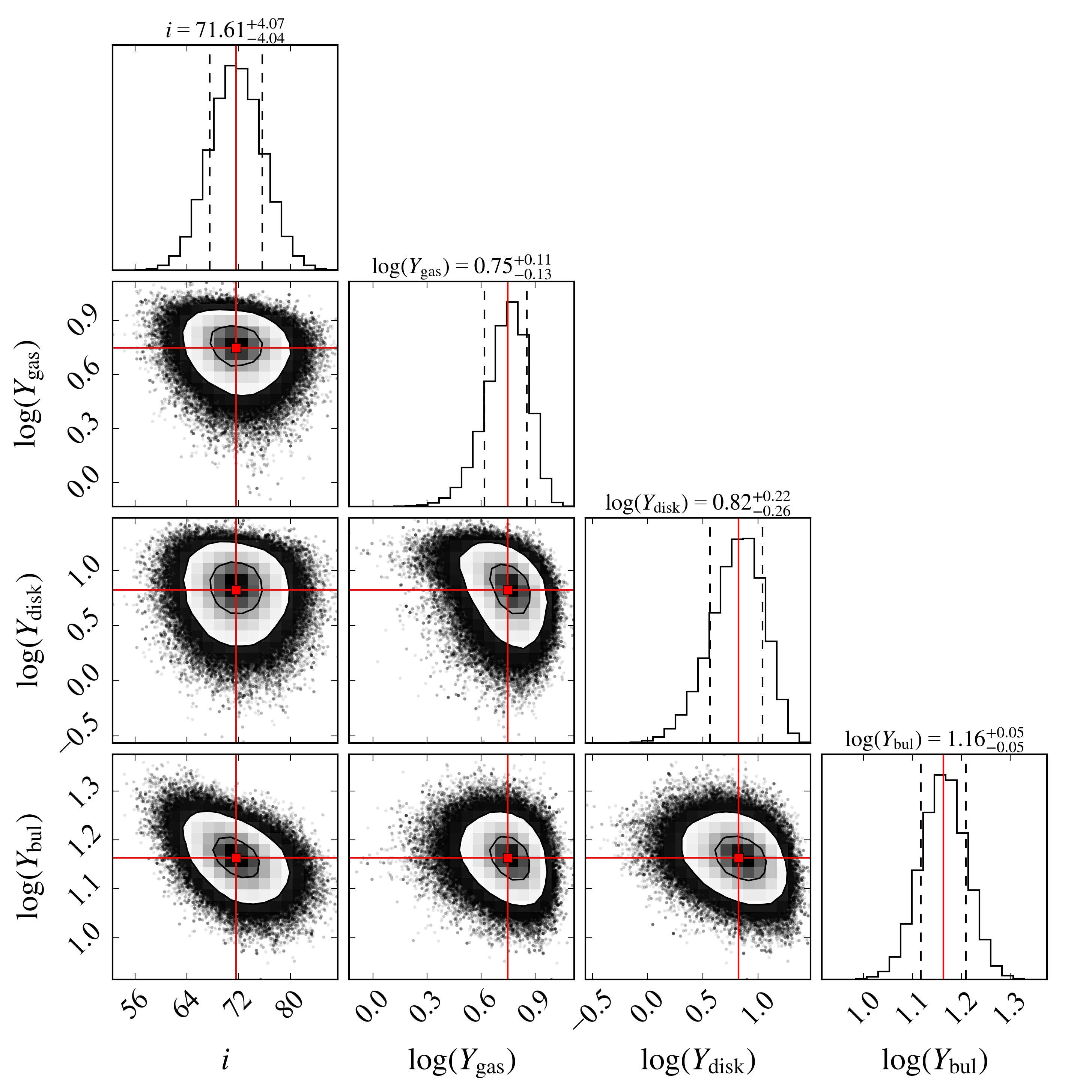}
\includegraphics[width=0.49\textwidth]{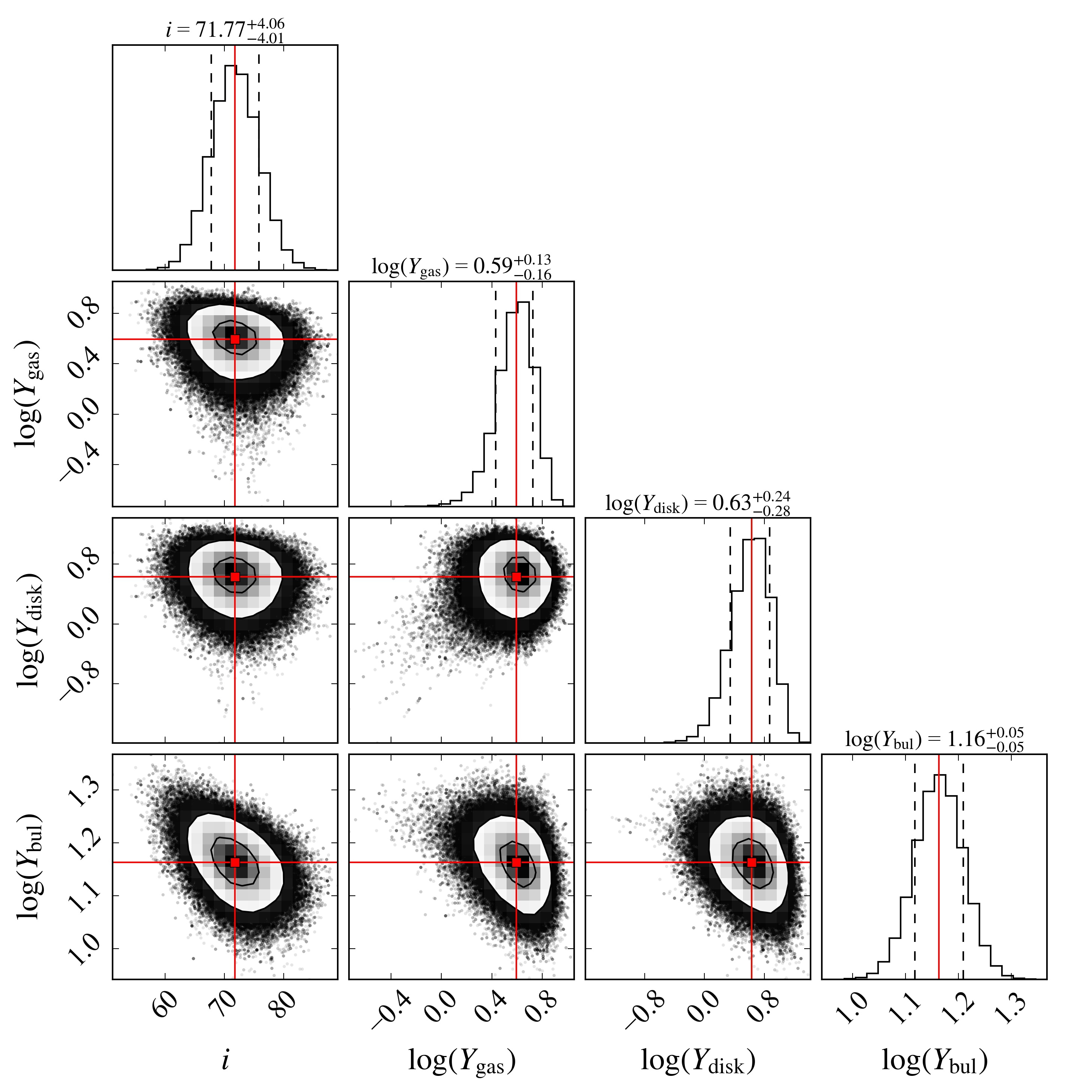}
\includegraphics[width=0.72\textwidth]{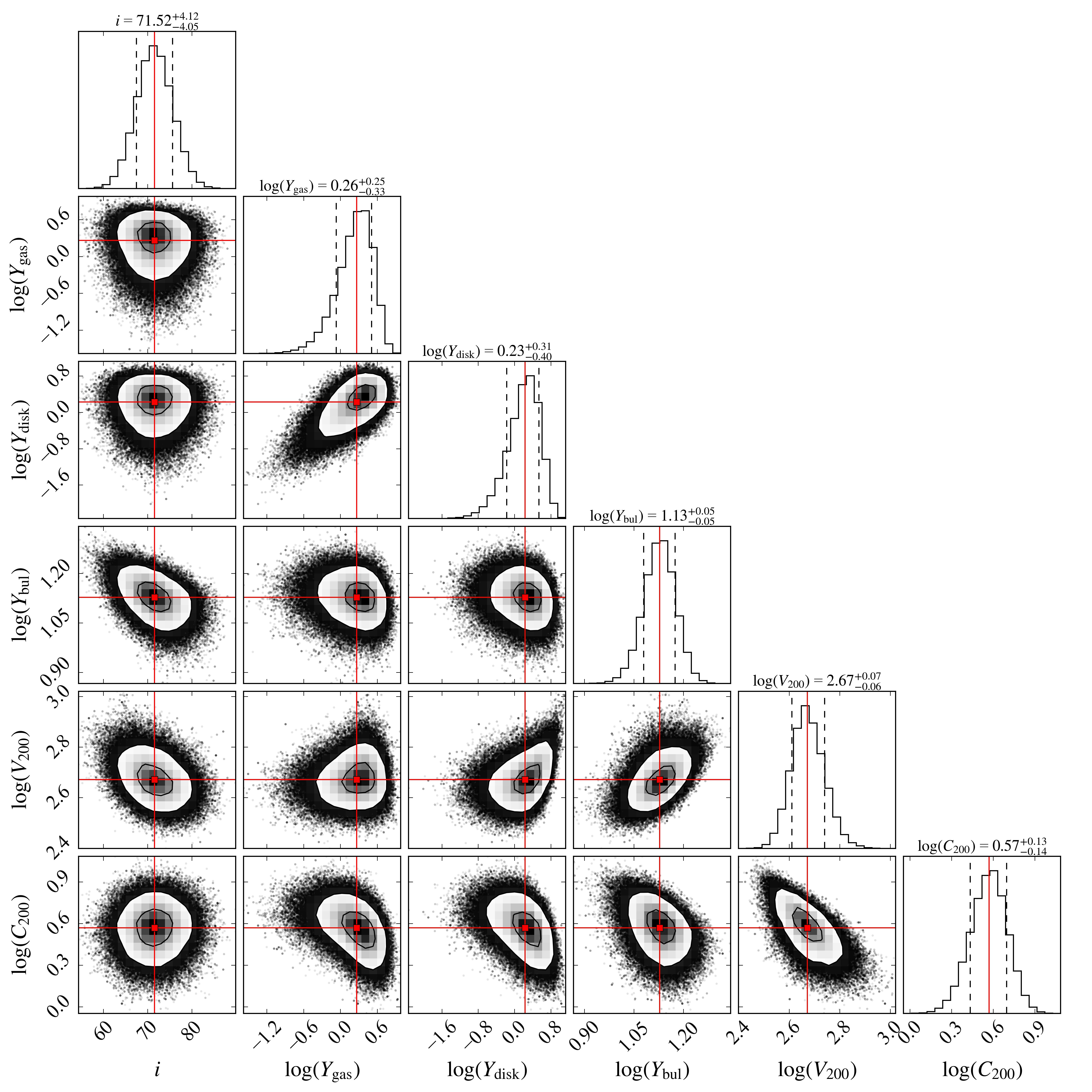}
\caption{Same as Fig.\,\ref{fig:CornerBird1} but for zC-488879.}
\label{fig:CornerBird2}
\end{figure*}

Figures\,\ref{fig:CornerBird1} and \ref{fig:CornerBird2} present ``corner plots'' from MCMC fits to the rotation curves (see Sect.\,\ref{sec:massmodels}), using the \texttt{corner.py} package \citep{Corner2016}. In general, the posterior probability distributions are well-behaved and show clear peaks, indicating that the fitting quantities are well measured.

\end{appendix}

\end{document}